\documentclass[manuscript]{aastex}

\usepackage{color}
\usepackage[normalem]{ulem}

\shorttitle{SN~Ia as sites for $p$-process }
\shortauthors{Travaglio et al.}

\begin{document}


\title{Type Ia supernovae as sites of $p$-process: two-dimensional models coupled to nucleosynthesis}

\author{C. Travaglio\altaffilmark{1}}
\affil{INAF - Astronomical Observatory Turin, Italy}
\affil{B2FH Association - Turin, Italy}
\email{travaglio@oato.inaf.it, claudia.travaglio@b2fh.org}

\author{F. K. R{\"o}pke\altaffilmark{2}}
\affil{Universit{\"a}t W{\"u}rzburg, Am Hubland, D-97074 W{\"u}rzburg, Germany}
\affil{Max-Planck-Institut f\"ur Astrophysik, Karl-Schwarzschild-Str.~1, D-85748 Garching bei M\"unchen, Germany}
\email{fritz@mpa-garching.mpg.de}

\author{R. Gallino\altaffilmark{3}}
\affil{Dipartimento di Fisica Generale, Universit\'a di Torino, Italy}
\affil{B2FH Association - Turin, Italy}
\affil{INAF - Astronomical Observatory Teramo, Italy}

\and

\author{W. Hillebrandt\altaffilmark{2}}
\affil{Max-Planck-Institut f\"ur Astrophysik, Karl-Schwarzschild-Str.~1, D-85748 Garching bei M\"unchen, Germany}

\begin{abstract}
Beyond Fe, there is a class of 35 proton-rich nuclides, between
$^{74}$Se and $^{196}$Hg, called $p$-nuclei.  They are bypassed by the
$s$ and $r$ neutron capture processes, and are typically 10$-$1000
times less abundant than the $s$- and/or $r$-isotopes in the Solar
System.  The bulk of $p$ isotopes is created in the 'gamma processes' by
sequences of photodisintegrations and beta decays in explosive
conditions in both core collapse supernovae (SNII) and in Type Ia
supernovae (SN~Ia). SNII contribute to the production of $p$-nuclei
through explosive neon and oxygen burning.  However, the major problem
in SNII ejecta is a general underproduction of the light $p$-nuclei,
for A $<$ 120.  We explore SNe~Ia as $p$-process sites in the framework
of two-dimensional SN~Ia delayed detonation model as well as pure
deflagration models. The WD precursor is assumed to have reached the
Chandrasekhar mass in a binary system by mass accretion from a
giant/main sequence companion.  We use enhanced $s$-seed
distributions, with seeds directly obtained from a sequence of thermal
pulse instabilities both in the AGB phase and in the accreted
material.  We apply the tracer-particle method to reconstruct the
nucleosynthesis by the thermal histories of Lagrangian particles,
passively advected in the hydrodynamic calculations. For each particle
we follow the explosive nucleosynthesis with a detailed nuclear
reaction network for all isotopes up to $^{209}$Bi. We
select tracers within the typical temperature range for $p$-process
production, 1.5$-$3.7 10$^{9}$K, and analyse in detail their behavior,
exploring the influence of different $s$-process distributions on the
$p$-process nucleosynthesis. In addition, we discuss the sensitivity
of $p$-process production to parameters of the explosion mechanism,
taking into account the consequences on Fe and alpha elements. We find
that SNe~Ia can produce a large amount of $p$-nuclei, both the light
$p$-nuclei below $A$=120 and the heavy-$p$ nuclei, at quite flat
average production factors, tightly related to the $s$-process seed
distribution. For the first time, we find a stellar source able to
produce both, light and heavy $p$-nuclei almost at the same level as
$^{56}$Fe, including the very debated neutron magic $^{92,94}$Mo and
$^{96,98}$Ru.  We also find that there is an important contribution
from $p$-process nucleosynthesis to the $s$-only nuclei $^{80}$Kr,
$^{86}$Sr, to the neutron magic $^{90}$Zr, and to the neutron-rich
$^{96}$Zr.  Finally, we investigate the metallicity effect on
$p$-process production in our models.  Starting with different
$s$-process seed distributions, for two metallicities $Z$=0.02 and
$Z$=0.001, running two-dimensional SNe~Ia models with different initial
composition, we estimate that SNe~Ia can contribute to, at least, 50\% 
of the solar $p$-process composition. A more detailed analysis 
of the role of SNe~Ia in Galactic chemical evolution of $p$-nuclei is
in preparation.
\end{abstract}


\section{Introduction}

Among the nuclei heavier than $^{56}$Fe, there is a class of 35
nuclides called $p$-nuclei, which are typically 10$-$1000 times less
abundant than the $s$- or $r$-isotopes in the Solar System. They
cannot be synthesized by neutron-capture processes since they are
located on the neutron-deficient side of the valley of
$\beta$-stability.  The astrophysical origin of $p$-nuclei has been
studied for 50 years, starting from the pioneering works by
Cameron~(1957) and Burbidge et al.~(1957). They suggested that a
combination of proton captures and photodissociations of $s$- and
$r$-seed nuclei could produce $p$-nuclei at temperature between 2 and
3~10$^9$ K. About 20 years later, different attempts were made to
explain the synthesis of all $p$-nuclei in one astrophysical site
(Audouze \& Truran~1975; Arnould~1976; Woosley \& Howard~1978). Those
authors suggested that the largest fraction of $p$-isotopes in the solar
system should be created by photodisintegration (the so-called
`$\gamma$-process') reactions operating upon a distribution of
$s$-process seeds synthesized in the earlier evolutionary stages of
the progenitor.  They suggested Type II supernovae (hereafter SNII) as
the astrophysical site of the $p$-process, and demonstrated that the
$p$-process is extremely sensitive to temperature and timescales.
Detailed calculations of $p$-process nucleosynthesis in SNII have been
performed by many authors, e.g. Woosley \& Howard~(1990), Rayet et
al.~(1900, 1995), Rauscher et al.~(2002), Hayakawa et al. ~(2006,
2008), Farouqi et al.~(2009). In these studies, core-collapse supernovae have been considered
best candidates to reproduce the solar abundances of the bulk of
the $p$-isotopes. However, it has been shown that the `gamma-process'
scenario suffers from a strong underproduction of the most abundant
$p$-isotopes, $^{92,94}$Mo (see e.g. Fisker et al.~2009) and
$^{96,98}$Ru.  For these nuclei, alternative processes and sites have
been proposed, either strong neutrino fluxes in the deepest layers of
SNII ejecta (Fr\"olich et al.~2006), or rapid proton-captures in
proton-rich, hot matter accreted onto the surface of a neutron star
(e.g. Schatz et al.~(2001)), ${\nu}p$-process in neutrino driven winds
of SNII (e.g. Fr\"olich et al.~2006; Pruet et al.~2006; Wanajo et
al.~2006, 2011a, 2011b). Woosley et al.~(1990) additionally introduced a
$\nu$-process to reproduce $^{138}$La and $^{180m}$Ta. Recent
multidimensional SNII models showed that the ejecta can become
proton-rich for several seconds (Fisher et al.~2010), and the
importance of the nucleosynthesis (including the $\nu$$p$-process) in
the reverse shock has been studied in detail (Wanajo et al.~2011a;
Roberts et al.~2010; Arcones \& Janka~2011).  Fujimoto et al.~(2007)
performed calculations of the composition of magnetically driven jets
ejected from a collapsar, based on magnetohydrodynamic
simulations of a rapidly rotating massive star during core
collapse. They found that not only the $r$-process successfully operates
in the jets (with a pattern inside the jets similar to that of the
$r$-elements in the solar system), but also $p$-nuclei are produced
without the need of $s$-seeds. Light $p$-nuclei, such as $^{74}$Se,
$^{78}$Kr, $^{84}$Sr, and $^{92}$Mo, are abundantly synthesized in the
jets, together with $^{113}$In, $^{115}$Sn, and $^{138}$La. They claim
that the amounts of $p$-nuclei in the ejecta are much larger than
those in core-collapse supernovae.  More recently a different
possibility to synthesize light $p$-nuclei has been presented by
Wanajo et al.~(2011a) and Arcones \& Montes~(2011). These authors
explored the possibility of synthesis of $p$-nuclei in proton-rich
winds of SNII.  Other possible scenarios for the production of
$p$-nuclei have been proposed by Iwamoto et al.~(2005) considering
hypernovae, by Fujimoto et al.~(2003) suggesting accretion disks around
compact objects, and by Nishimura et al.~(2006) investigating jet-like
explosions.

The $p$-process has also been suggested to occur in the outermost
layers of SNe~Ia, based on a delayed-detonation
model (Howard \& Meyer~(1993)), He-detonation models for a
sub-Chandrasekhar mass white dwarf (WD) (Goriely et al.~(2005);
Arnould \& Goriely~(2006)), and by Kusakabe et al.~(2005,
2011) based on the W7 carbon deflagration model (Nomoto et
al.~1984). All these authors considered both solar abundances as seeds
for the $p$-process, as well as $s$-enhanced seeds.  Goriely et
al.~(2002), presenting 1D He-detonating, sub-Chandrasekhar mass CO-WD
models, and Goriely et al.~(2005), presenting 3D explosion models,
concluded that sub-Chandrasekhar He-detonation models are not an
efficient site for the synthesis of $p$-nuclei.

Kusakabe et al.~(2005, 2011) analyzed the effects of different $s$-seed
distributions on $p$-production. They derived the $s$-seed
distribution by assuming an exponential distribution of neutron
exposures with two choices of the mean exposure $\tau_0$ = 0.30
mb$^{-1}$, which best reproduces the main component in the Solar
System (Arlandini et al.~1999, classical model), or $\tau_0$ = 0.15
mb$^{-1}$, which gives rise to a $s$-process distribution decreasing
with increasing atomic mass number $A$. The accuracy of the treatment
of the outer zones of the SN~Ia is fundamental for the $p$-nuclei
production, and the sparse zoning in the outermost layers of the W7
model has to be taken into account for the thorough of $p$-process
nucleosynthesis calculations. As we will describe in the paper, our
multi-dimensional SN~Ia models can follow even the outermost parts of
the star quite accurately .

We have calculated $p$-process nucleosynthesis with
high-resolution two-dimensional hydrodynamic models of SN~Ia
considering both a pure deflagration (similar to R\"opke et al.~2006)
and delayed detonations of different explosion strengths (similar to
those presented in Kasen, R\"opke \& Woosley~2009).  We also have
calculated $p$-process nucleosynthesis for SN~Ia of metallicity lower
than solar. The adopted SN~Ia models are detailed in Section~2.  The
tracer particles method to calculate the nucleosynthesis in multi-D
simulations (see e.g. Travaglio et al.~2004, Maeda et al.~2010),
together with the nuclear reaction network is described in Section~3.
We consider different $s$-process seed
distributions, as detailed in Section~4.  In Section~5 we show an
analysis of the production mechanisms of all $p$-nuclei.  The results
for the $p$-process production, depending on the different SN~Ia model
adopted and on the $s$-process distribution tested, are discussed in
Section~6.  Finally, conclusions and work in progress are described in
Section~7.

\section{Type Ia supernova models}

SNe~Ia are associated with thermonuclear explosions of white dwarf (WD)
stars that are composed of carbon and oxygen. A favored scenario is
that the explosion is triggered once the WD approaces the Chandrasekhar
mass. In the \emph{single-degenerate scenario} which we refer to here,
this happens due to accretion of material from a main-sequence or
evolved companion star in a binary system. We follow the explosion
phase of Chandrasekhar-mass WD in two-dimensional hydrodynamic
simulations. The numerical method employed has been described
in detail by Reinecke et al.~(1999); R{\"o}pke \& Niemeyer~(2007);
R{\"o}pke \& Schmidt~(2009); R{\"o}pke~(2005).

For the explosion, a number of scenarios has been suggested (see
Hillebrandt \& Niemeyer 2000 for a review). Here, we focus on pure
deflagrations and delayed detonations. In the former, the burning
front propagates subsonically throughout the explosion. It is subject
to instabilities that drive turbulence due to which the flame is
strongly accelerated beyond its laminar burning speed. However, the pure
deflagration model is not considered a candidate for most of the
normal SN~Ia as it falls short of reproducing the necessary explosion
energies and $^{56}$Ni masses (R{\"o}pke et al.~2007) although it may be an
explanation for the peculiar sub-class of 2002cx-like SN~Ia (Phillips
et al.~2007).  A transition of the flame propagation mode from
subsonic deflagration to supersonic detonation in a late stage of the
burning constitutes the delayed detonation model (Khokhlov~1991). The
necessary deflagration-to-detonation transition (DDT) may arise from
strong turbulence in the so-called distributed burning regime
(e.g. R{\"o}pke~2007; Woosley~2007; Woosley et al.~2009), but rigorous
evidence for its occurence in SN~Ia is still lacking and details of its
physical mechanism are uncertain. Nonetheless, delayed detonations
successfully reproduce main observables of normal SN~Ia (Mazzali et
al.~2007; Kasen et al.~2009) and are therefore considered a standard
model for these events.

Our hydrodynamic simulations do not follow the evolution of the WD
towards ignition, but start at the onset of the explosion. Here, we
ignite a cold ($T = 5.0 \times 10^5 \, \mathrm{K}$) Chandrasekhar-mass
WD in hydrostatic equilibrium with a chemical composition that results
from three different evolutionary phases of the progenitor,
i.e.\ central He-burning, He-shell burning during the early AGB phase,
and He-shell burning during the thermally pulsing AGB phase
(Dom\'inguez et al.~2001).  According to Piro \& Bildsten~(2008) and
Jackson et al.~(2010), we assume that as soon as the accreting WD
approaches the Chandrasekhar mass carbon-burning starts to occur in
the core. The energy released by this burning drives convection (the
so called {\it simmering phase}) that lasts 
for about 1000 years before the explosion. The extension of this 
convective zone is not well determined, and the most external zones 
beyond 1.2 $M_\odot$ may well remain unmixed (Piro \& Bildsten~2008). 
Note that the zone where the bulk of the $p$-isotopes is produced is 
the most external one, with an extension of the order of $\sim$0.1 $M_\odot$.

For our DDT-a model we use a CO-WD structure presented by Dom\'inguez et
al.~(2001) (Table~1), with $Z =$ 0.02 and a progenitor mass of $M =$
1.5$M_\odot$. The sensitivity to the different CO-WD
structure (obtained evolving models with different main-sequence mass)
and the uncertainties in the extension of the simmering phase, suggest
that a uniform C/O ratio with 50\% mixture for the burning can be a
good approximation.  Concerning the initial Y$_e$, two different
setups are used, the first one, marked with \textbf{a}, assumes $Y_e =
0.499$, i.e.\ close to solar metallicity\footnote{Note that in our
hydrodynamic simulations, $Y_e$ is treated as an independent
parameter and does not reflect the actual chemical composition of
the WD material as modeled here.} $Z_\odot$.  Setup \textbf{b}
assumes $Y_e = 0.4995$ (corresponding to $\sim 1/20\, Z_\odot$).  All
simulations were performed with $512 \times 1024$ non-uniform moving
computational grid cells adopting the nested-grid technique of
R{\"o}pke \& Hillebrandt~(2005).

In all models discussed in this paper, the thermonuclear burning front
is ignited in multiple sparks (R{\"o}pke et al.~2006) near the center
of the WD. Specifically, we chose the setup of model DD2D\_iso\_05 of
Kasen et al.~(2009) (90 ignition kernels of $6 \, \mathrm{km}$ radius
randomly placed in radial direction according to a Gaussian
distribution with a width of $150 \, \mathrm{km}$) for all but one
model, where we employ the ignition kernel distribution of model
DD2D\_iso\_01 (20 ignition kernels of the same radius placed according
to a Gaussian radial distribution with a standard deviation of $150 \,
\mathrm{km}$).

The chosen WD setups and the explosion scenarios considered result
in five different hydrodynamic explosion models:
\begin{itemize}
  \item \textbf{DEF-a}, a pure deflagration model assuming the WD setup \textbf{a},
  \item \textbf{DDT-a}, a delayed detonation model assuming the WD setup
    \textbf{a} and the deflagration-to-detonation criterion dc2 of Kasen
    et al.~(2009). Detonations are triggered at any point on the
    deflagration flame where a Karlovitz number of $\mathrm{Ka} = 250$
    is reached while the fuel density is in the range $0.6 \lesssim
    \rho_\mathrm{fuel}/(10^7\, \mathrm{g} \, \mathrm{cm}^{-3}) \lesssim
    1.2$,
  \item \textbf{DDT-b}, a delayed detonation model assuming the WD setup
    \textbf{b} and the same deflagration-to-detonation criterion dc2 of
    Kasen et al.~(2009),
  \item \textbf{DDTw-a}, a delayed detonation model assuming the WD
    setup \textbf{a} and the deflagration-to-detonation criterion dc4 of
    Kasen et al.~(2009), i.e.\ detonations are this time triggered at
    any point on the deflagration flame where a Karlovitz number of
    $\mathrm{Ka} = 1500$ is reached while the fuel density is in the
    range $0.6 \lesssim \rho_\mathrm{fuel}/(10^7\, \mathrm{g} \,
    \mathrm{cm}^{-3}) \lesssim 1.2$, 
  \item \textbf{DDTs-a}, a delayed detonation model assuming the WD
    setup \textbf{a}, the ignition kernel distribution DD2D\_iso\_01 and
    the deflagration-to-detonation criterion dc2 of Kasen et al.~(2009).
\end{itemize}

The results of the hydrodynamic explosion simulations are summarized
in Table~1. Those of the delayed detonation models are not identical
(but reasonably close to) the results of the corresponding models
presented in Kasen et al.~(2009) where different initial WD models
were used. The pure-deflagration model produces little $^{56}$Ni and the
explosion energy is particularly low because -- as characteristic for
pure deflagration models -- burning ceases quickly after nuclear
statistical equilibrium (NSE) is no longer reached in the ashes and
only about a tenth of a solar mass of intermediate-mass elements is
produced. More than half of the WD mass remains unburned. Since the
iron group nuclei are synthesized at rather high densities,
neutronization is efficient and a large fraction of the iron group
nuclei are Fe and Ni stable isotopes rather than $^{56}$Ni. As
expected, the DD2D\_iso\_05 ignition kernel setup in combination with
the dc2 DDT criterion leads to an explosion strength
($E_\mathrm{kin}^\mathrm{asym} \sim 1.3 \times 10^{51} \,
\mathrm{erg}$) and a $^{56}$Ni mass ($\sim 0.5 \, M_\odot$) that are
typical for ``normal'' SN~Ia and the effect of the two WD models ({\bf
  a} vs.\ {\bf b}) is of secondary importance for the global explosion
characteristics.  The two additional models, DDTw-a and DDTs-a were
computed to explore the range of explosion strengths that is observed
for normal SN~Ia.  With $^{56}$Ni masses of $0.301 \, M_\odot$ and
$0.951 \, M_\odot$ they capture the extreme limits of the range (or
perhaps even exceed them slightly).

The multidimensional SN~Ia simulations described above assume instant
burning of the initial C+O material once crossed by a deflagration or
detonation front (which is represented as a discontinuity applying the
level-set method). The microphysical details of the burning are not
resolved. Instead, the fuel material is converted to an ash
composition according to the fuel density ahead of the front. The ash
material is modeled by a mixture of $^{56}$Ni and $\alpha$-particles
representing NSE that is dynamically adjusted according to the
thermodynamic conditions during the simulation and by a hypothetical
(but representative) intermediate-mass nucleus ($A=30$,
$E_\mathrm{bind} = 6.8266 \times 10^{18} \, \mathrm{erg}\,
\mathrm{g}^{-1}$). At the lowest fuel densities above the burning
threshold, burning from carbon to oxygen (which also mimics burning to
neon) is included. The neutronization of the material is treated
independently of this composition by advecting $Y_e$ as a passive
scalar and taking into account its evolution due to electron capture
reactions in the NSE material. This coarse treatment of nuclear
reactions is sufficient to account with sufficient precision for the
energy release that drives the explosion dynamics. A more detailed
treatment, however, is necessary to analyze the specific
nucleosynthetic processes which are the focus of our present work.  To
this end, a number of Lagrangian tracer particles that record
thermodynamic trajectories is passively advected with the hydrodynamic
flow in the explosion simulation. On that basis, a post-processing
step is performed using an extended network up to $^{209}$Bi as 
described in the following Section. The masses of the species in
the ejecta quoted above (and shown in Table~1) are therefore only an
estimate. More reliable values are obtained in the nucleosynthetic
postprocessing step.

\section{Nucleosynthesis in multi-D SN~Ia}

The multidimensional hydrodynamic scheme we use follows the explosion 
by means of an Eulerian grid. In order to follow over time the temperature
and density evolution of the fluid we introduce a Lagrangian component
in form of tracer particles. During the hydrodynamical simulation,
they are advected by the flow, recording the $T$ and $\rho$ history
along their paths. The nuclear post-processing calculations are then
performed for each particle separately.  The tracer particles method
for nucleosynthesis calculations for core-collapse SNe has been
introduced first by Nagataki et al.~(1997), and for SN~Ia by Travaglio
et al.~(2004, 2005).

For the present work we use 51200 tracer particles, uniformly distributed 
in mass coordinate. Each tracer represents the same mass
of $\simeq$2.73 $\times$ 10$^{-5} M_\odot$ (= $M_\mathrm{WD}$/51200).  The
distribution of the tracer particles for our standard DDT-a model is
shown in Figure~1 in several snapshots illustrating the evolution.  We
compare the density distribution obtained in the hydrodynamical
simulation with the distribution of the tracers.  Different colours
are used for different ranges of peak temperature of the tracers
(i.e. $T_\mathrm{peak}$, the maximum $T$ reached by a tracer
throughout the entire explosion).  The tracers marked in {\it black}
are responsible for the Fe-group production. With their maximum
temperature above $T_9 \ge$ 7 (where $T_9$ is the temperature in
units of 10$^9$ K) they reach NSE and most of the nucleosynthesis goes
to $^{56}$Ni or Fe-group nuclei. The {\it grey tracers} are instead
the main producers of the lighter $\alpha$-isotopes. In {\it red},
{\it green} and {\it blue} we plotted the tracers with $T_9 \leq$ 3.7,
the main contributors to the $p$-process nuclei.  Close to the surface
of the white dwarf, the peak temperature reached during the explosive
phase is not high enough for the nuclei to attain nuclear statistic
equilibrium condition, although significant transmutation of heavy
elements into $p$-process nuclei occurs.  The three ranges within this
group, 3.0$< T_9 \leq$ 3.7 ({\it red}), 2.4$< T_9 \leq$ 3.0 ({\it
  green}), and 1.5$\le T_9 \leq$ 2.4 ({\it blue}), are connected to
three different behaviors we identify for the production of $p$-nuclei
(see Section~5).  This figure shows that there are tracers with low
$T_\mathrm{peak}$ ({\it red}) in the inner part of the star, resulting
from low-density burning in the deflagration regime.  However, the
bulk of the {\it red} tracers is located in the outermost part of the
star, together with {\it green} and {\it blue} tracers, where in our
models most of the mass has been accreted from a companion and is burned 
in the delayed detonation models at low densities in the
detonation phase.

Recently Seitenzahl et al.~(2010) demonstrated that in 2D SN~Ia
simulations with 80$^2$ tracers, all isotopes up to Mo with abundances
higher than $~$10$^{-5}$ are reproduced with an accuracy of 5$\%$
(with the exception of $^{20}$Ne).  We also performed a resolution
study, which are discussed in Section~6.5.  
The nucleosynthesis of the main species obtained with the tracer particles method
(i.e. $^{56}$Fe, $^{12}$C, $^{16}$O as well as the mass involved for the
$p$-process nucleosynthesis) are summarized in Table~2. In the
last column we report for comparison the abundances for
W7 SN~Ia model (Iwamoto et al.~1999).

\subsection{$p$-process nucleosynthesis}

The $p$-process nucleosynthesis is calculated using a nuclear network
with 1024 species from neutron and proton up to $^{209}$Bi combined
with neutron, proton and $\alpha$ induced reactions and their
inverse. The code used for this work was originally developed and
presented by Thielemann et al.~(1996).  We use the nuclear reaction
rates based on the experimental values and the Hauser-Feshbach
statistical model NON-SMOKER (Rauscher \&
Thielemann~2000). Theoretical and experimental electron capture and
$\beta$-decay rates are from Langanke \& Mart\'inez-Pinedo~(2000).

The currently favoured production mechanism for those isotopes is
photodisintegration of intermediate and heavy nuclides.  The
$\gamma$-process becomes effective during the explosive O-burning
phase, at $T_9 \ge 2.5$, and starts with the photodisintegration of
stable seed nuclei. During the photodisintegration period, proton,
neutron and $\alpha$-emission channels ($\gamma$,n), ($\gamma$,p), and
($\gamma$, $\alpha$) compete with each other and with $\beta$-decays
of nuclides far from stability.

In Figure~2 we plot the abundance variations for proton, neutron and
$^{4}$He as a function of time for selected tracer particles, at
different peak temperatures, in the range between $T_9$=1.5 and
$T_9$=3.7. Comparing with the analogous plot presented in Figure~3 of
Kusakabe et al.~(2011), we find similar abundances for p, n and
$^{4}$He, but the timescales in our explosion model are longer. A
difference to notice, however, is the double peak for some of the
selected tracers. This can be attributed to the fact that our model
involves both a deflagration phase and a detonation phase, whereas
Kusakabe et al.~(2011) employ a pure deflagration model.
  
In Table~3 we list the 35 $p$-isotopes with their relative abundance
(in per cent) to the respective elements in the Solar System.  In
Table~4 we list additional isotopes for which we get an
important $p$-contribution -- among them the $s$-only isotopes
$^{80}$Kr and $^{86}$Sr and the neutron magic $^{90}$Zr. We also
include the neutron-rich isotope $^{96}$Zr, which is substantially
produced by neutron capture via the $^{22}$Ne($\alpha$, n)$^{25}$Mg
chain during the carbon-burning phase (see Sections~5 and ~6.2 for a
more detailed discussion). In the same Table, we additionally list the
neutron magic isotopes $^{86}$Kr, $^{87}$Rb, $^{88}$Sr and $^{89}$Y, 
which are very abundant in the $s$-seed and they are not affected by
photodisintegrations (see Section~6.2). Consequently these isotopes
have to be considered as relics of the $s$-process seeds.

\section{$s$-process seeds}

The $p$-process nucleosynthesis occurs in SN~Ia only if there is an
$s$-process enrichment, and therefore it is essential to determine the
$s$-process enrichment in the exploding WD.  In the single-degenerate
progenitor model which we assume here, there are two sources of
$s$-enrichment: (1) during the Asymptotic Giant Branch phase leading
to the formation of the WD, thermal pulses occur (TP-AGB phase) in
which $s$-isotopes are produced (see e.g. Dom\'inguez et al.~2001;
Straniero et al.~2006), (2) thermal pulses during the accretion phase
enrich the matter accumulated onto the WD (Iben~1981; Iben \&
Tutukov~1991; Howard \& Meyer~1993; Kusakabe et al.~2011). The
$s$-enrichment of the WD in a layer of $\sim$ 0.1 $M_\odot$ deriving
from the past AGB history, prior the accretion phase, is convectively
mixed into the WD (see discussion of the simmering phase in Piro \&
Bildsten~2008; Piro \& Chang~2008; Chamulak et al.~2008). 
This dilutes the $s$-seeds so that their abundances
are too low for producing significant yields of $p$-isotopes.  In
addition, $s$-process nucleosynthesis can occur in the H-rich matter
accreted by the CO-WD, due to recurrent He-flashes (Iben~1981), with
neutrons mainly produced by the $^{13}$C($\alpha$,n)$^{16}$O
reaction. 

We investigate the influence of different $s$-process abundance
distributions for $Z = 0.02$ and $Z = 0.001$ (see Figs.~3 and ~4) as
seeds of $p$-process nucleosynthesis, These distributions are obtained
from $s$-process nucleosynthesis calculations with a post-process
method (Gallino et al.~1998; Bisterzo et al.~2010). In the AGB
scenario, there is a general consensus (Gallino et al.~1998 and
references therein) that the main neutron source is the reaction
$^{13}$C($\alpha$,n)$^{16}$O.  In order to activate it, partial mixing
of protons from the envelope down into the C-rich layers is required
(physical causes of this mixing are discussed by many authors,
e.g. Hollowell \& Iben~1988; Herwig et al.~1997; Goriely \& Mowlawi~2000; 
Langer et al.~1999; Herwig et al.~2003; Denissenkov \& Tout~2003; 
Cristallo et al.~2009). However, the mass involved and the
profile of the $^{13}$C-pocket have still to be considered as free
parameters, given the difficulty of a realistic treatment of the
hydrodynamical behavior at the H/He discontinuity. A series
of constraints have been obtained by comparing spectroscopic
abundances in $s$-enriched stars at different metallicities with AGB
model predictions (see e.g., Busso et al.~2001; Bisterzo et al.~2010).
The spread in the $s$-process yields at each metallicity has
been modeled parametrically by varying the $^{13}$C concentration in
the pocket from 0 up to a factor of 2 times the {\it standard} value
of $\sim 4 \times 10^{-6} M_\odot$ of $^{13}$C (Gallino et al.~1998,
ST case), and is indicated in Figs.~3 and ~4 as ST, ST$\times$2
(standard $^{13}$C-pocket multiplied by a factor of 2), ST/2 (standard
$^{13}$C-pocket divided by a factor of 2), etc.  For this work we
varied the $^{13}$C-pocket concentration in order to obtain a flat
$s$-seeds distribution, or a non-flat $s$-seeds distribution peaked 
to the lighter or to the heavier $s$-nuclei
(see Figs.~3 and ~4), for both $Z$=0.02 and $Z$=0.001.

In Figure~3 we show for two metallicities a typical flat $s$-process
distribution (i.e. the isotopes produced mainly by $s$-process
nucleosynthesis have all the same overabundance with respect to the
solar abundances). The {\it upper panel} is a flat $s$-process
distribution similar to the one presented by Arlandini et al.~(1999, 
'stellar model') as a best fit to the solar main-component.  In Arlandini 
et al.~(1999), a full stellar evolutionary model is followed along
the TP-AGB phase, using an updated network of neutron captures and
beta decay rates (Takahashi \& Yokoi~1987; Bao et al.~2000) updated 
with more recent experimental determinations in the KADONIS database 
(Dillmann et al.~2006).

The larger number of thermal pulses that in principle can be realized
during mass accretion, more easily allows the $s$-nuclei to achieve an
asymptotic distribution. The difference with respect to Kusakabe et
al.~(2011) case B, for a flat distribution in the heavy $s$-only
isotope overabundances (about 7000 as compared to our 2000),
may be ascribed to different reasons. First, their choice of the
neutron capture MACS (Maxwellian Averaged Cross Section) is based on
the use of the current 30 keV data. At 30 keV, the MACS of $^{56}$Fe, the major seed
for the $s$-process, is 11.7$\pm$0.4 mb (Bao et al.~2000). However
its temperature dependence strongly departs from the usual 1/$v$
rule. Actually, the MACS is almost flat between 100 keV and 7 keV
(Beer, Voss \& Winters~1992, see their Figure~3). In AGB stellar modes, 
the major neutron source is the $^{13}$C($\alpha$,n)$^{16}$O reaction. Neutrons are
released at about 8 keV. Using a value of 30 keV value instead is
equivalent to double the effective MACS of $^{56}$Fe. Second, Kusakabe et al. derived
the $s$-process distribution using a simplified exponential
distribution of neutron exposures. This is in agreement to the
classical analysis of the $s$-process, using for the mean neutron
exposure parameter the value $\tau_0$ = 0.30 mb$^{-1}$. The result is
a flat distribution for the $s$-only isotopes beyond $A \sim$ 90
(Ulrich~1973).

The classical analysis operates at fixed temperature and neutron
density assumed parametrically and using an unbranched s-process flow,
not accounting for the much more complex astrophysical situation
occurring during recurrent thermal pulses in AGB stars. There, two
neutron sources operate at different thermal conditions. The major
neutron exposure is activated at $kT$ $\sim$ 8 keV by the
$^{13}$C($\alpha$,n)$^{16}$O reaction in radiative conditions between
two subsequent convective thermal instabilities in the top layers of
the He intershell (the so-called $^{13}$C-pocket). A second neutron
exposure results from the marginal activation of the
$^{22}$Ne($\alpha$,n)$^{25}$Mg reaction in the convective instability
at $kT$ $\sim$ 23 keV. Moreover, the classical analysis in principle
works only for an asymptotic distribution of neutron exposures over a
series of identical thermal instabilities and with constant overlap
factor between adjacent pulses. Besides the different thermal and
physical conditions, also the neutron density is far from being
constant.  For an exhaustive comparison see Gallino et al. (1998).
Besides, Kusakabe et al.~(2011, their case A) did not consider the neutron poison
effect of all light isotopes below $^{32}$S, which in particular
ignores $^{25}$Mg, the most important competitor of $^{56}$Fe in
neutron capture.

In Figure~4 we present different $s$-process distributions, used as
$s$-seeds for this work. In the {\it upper panel} we show a
$s$-distribution peaked at the light nuclei (with $A$ between $\sim$75
and $\sim$90). This resembles the pattern obtained from a classical
mean neutron exposure with $\tau_0$ = 0.15 mb$^{-1}$. Similar
conditions have been analyzed by Kusakabe et al.~(2011).  In the {\it
  middle panel} an analogous distribution is shown for $Z$ =
0.001. The {\it lower panel} of Figure~4 shows a distribution peaked
at heavy isotopes ($A \ge$ 150), with an important contribution at
$^{208}$Pb.  In Section~6 we will discuss in detail the
consequences for $p$-process nucleosynthesis using these different
distributions of $s$-seeds.

\section{Distribution of $p$-nuclei}

The production mechanism of the various $p$-nuclei can be understood
through an analysis of the corresponding nuclear flows. In
Figs.~5,~6,~7, and ~8 we plot the behavior of the different nuclei 
versus peak temperature for a solar metallicity case (i.e. DDT-a) 
using the flat $s$-process distribution shown in Figure~3 
({\it upper panel}).  In Figure~5 we show $^{12}$C, $^{16}$O, 
$^{20}$Ne, and $^{22}$Ne abundances as a function of $T_\mathrm{peak}$, 
for the tracers in the temperature range to produce $p$-nuclei.  
Starting from the cold outer layers of
the star, at $T_9 \simeq 1.4$ $^{22}$Ne burns through ($\alpha, n$)
reaction, becoming the most important source of neutrons. This happens
at about 0.6 sec after ignition of the SN~Ia.  At $T_9 \simeq$ 2
$^{12}$C burns mainly via ($^{12}$C,$^{12}$C) channels making
$^{23}$Na, $^{20}$Ne, $\alpha$ and $p$-nuclei. At $T_9 \simeq2.6$
photodisintegration of $^{20}$Ne via
$^{20}$Ne($\gamma$,$\alpha$)$^{16}$O becomes efficient which thus
increasing the amount of available $^{16}$O.  At $T_9 \simeq3.2$
photodisintegration of $^{16}$O becomes efficient. The range of $T_9$
chosen for Figure~5 is related to the $T_9$ range where the
$p$-process nucleosynthesis occurs.

In Figure~6 and Figure~7 we show the final abundances of all the 35
$p$-isotopes listed in Table~3 as a function of peak $T$.  In Figure~8
we plot the abundance behavior of the isotopes listed in Table~4,
i.e. the $s$-only $^{80}$Kr and $^{86}$Sr ({\it upper panel}),
$^{86}$Kr, $^{87}$Rb, $^{88}$Sr, $^{89}$Y ({\it middle panel}) to be
considered as relics of $s$-process seeds with a small contribution
from $^{22}$Ne burning, and the two special $^i$Zr isotopes $^{90}$Zr and
$^{96}$Zr ({\it lower panel}).

When $T_9 \simeq 2.2$, protons are released by burning of $^{12}$C, we
find that the first $p$-isotope produced at the lowest $T_9$ is
$^{180m}$Ta.  A behavior similar to $^{180m}$Ta is observed for
$^{184}$Os, but at somewhat higher temperatures ($2.4 \lesssim T_9
\lesssim 2.5$).

A bit further inside the star, where peak temperatures reach $T_9
\simeq$ 2.3 $-$ 2.4, the p isotopes $^{158}$Dy, $^{164}$Er, $^{180}$W,
$^{174}$Hf, $^{168}$Yb, $^{190}$Pt, $^{196}$Hg are produced. This
group of isotopes shows a second peak of production at higher $T$,
at $T_9 \simeq$ 2.7.  Also $^{152}$Gd belongs to this
group of $p$-isotopes, but here the second abundance peak at $T_9
\simeq$ 2.7 is much lower with respect to the first abundance peak at
$T_9 \simeq$ 2.4. Note that $^{152}$Gd is mainly produced by the
$s$-process, as recalled in Section~5.2.

When $T_9$ exceeds $\sim$ 3, ($\gamma, p$), ($\gamma,\alpha$) and
($\gamma$, n) become dominant and the lightest $p$-isotopes are
produced. $^{74}$Se, $^{78}$Kr, and $^{84}$Sr are synthesized in a
wide range of peak temperatures ($3.1 \lesssim T_9 \lesssim 3.6$). 
Also other light $p$-isotopes like $^{92,94}$Mo, $^{96,98}$Ru,
$^{102}$Pd, $^{106}$Cd, $^{112,114}$Sn are produced mostly at $T_9 \ge
$ 3.

In Figure~8 we show the abundance behavior as a function of peak
temperature for all isotopes reported in Table~4, i.e. those where we
find (as we will discussed in detail below) an important contribution
from $p$-process nucleosynthesis. From this Figure it is evident 
that $^{80}$Kr and $^{86}$Sr ({\it upper panel}), and $^{90}$Zr 
({\it lower panel}) are first destroyed by $^{22}$Ne burning, and
subsequently produced as $p$-nuclei in the $^{12}$C-burning phase, at
$T_9$ higher than 2. In contrast, the isotopes in the {\it middle panel},
i.e. $^{86}$Kr, $^{87}$Rb, $^{88}$Sr and $^{89}$Y, retain almost their
initial values (with a small increase for $^{86}$Kr and $^{87}$Rb)
during the $^{22}$Ne- and $^{12}$C-burning phases, while they are destroyed
by photodisintegration at $T_9 \ge$ 3. Finally, $^{96}$Zr ({\it lower panel}) 
is produced at $T_9 \simeq$ 1.6 due to $^{22}$Ne burning, in whose conditions 
the neutron density is high and the neutron capture channel on $^{95}$Zr is 
open. Note that the production yield of $^{96}$Zr directly depends on the 
uncertain theoretical maxwellian neutron capture cross section (MACS) of the unstable  
$^{95}$Zr. In our network we have adopted a factor of 2 lower than the one reported in 
the compilation of Bao et al.~(2000), taken as the average between the Bao et al.~(2000)
prescription and the much lower recent theoretical evaluation by the TALYS network of 
the Brussels datbase (http//www.astro.ulb.ac.de).
Several sources may contribute to the cosmic origin of the most neutron-rich
Zr isotope $^{96}$Zr (2.8\% of solar Zr), i.e. the weak $s$-process in massive stars, 
the main $s$-process in AGB stars of low and intermediate mass where the $^{22}$Ne 
neutron source is partially activated, as well as the weak $r$-process during SNII explosion. 
A further source that cannot be discarded anymore is now associated to the $p$-process in SN~Ia
explosions. Clearly, an experimental estimate of the MACS of the unstable $^{95}$Zr is of 
paramount importance.

\section{Results and discussion}

In this Section we present $p$-process nucleosynthesis results
obtained with different SN~Ia models, exploring different $s$-process
distributions, and investigating the metallicity effect on the
$p$-nuclei production.

\subsection{$p$-process in WD with AGB progenitor}

We present the results of the nucleosynthesis calculation for {\bf
  DDT-a}, i.e. a delayed-detonation model of a solar metallicity
Chandrasekhar-mass WD. We first consider the $s$-process enrichment of
the accreting WD as a result of its past AGB history. In this way the
mass of $\sim$ 0.1 $M_\odot$ enriched in $s$-process elements produced
during the AGB phase is spread out over the WD core (following Piro \&
Bildsten~2008). The resulting nucleosynthesis is presented in
Figure~9.  Nucleosynthesis calculations with the full network gave a
$^{56}$Fe yield of 0.584 $M_\odot$. This is consistent with what is
typically expected for a standard SN~Ia (Contardo et al.~2000;
Stritzinger et al.~2006). The detonation burns the WD almost
completely and only little $^{12}$C and $^{16}$O remain in the ejecta.
The average $p$-process enrichment we obtain is by a factor of about
50 to 100 below the $^{56}$Fe production. From the rough estimate
that SNe~Ia contribute 2/3 of the total Galactic $^{56}$Fe, we conclude
that this model cannot account for a significant fraction of
the solar $p$-nuclei.

\subsection{$p$-process in delayed detonation models with $s$-process
  in the accreted material}

The mass accreted in a close binary system from a hydrogen-rich
envelope of a companion onto the CO-WD can be enriched in $s$-process
material. As discussed by many authors in the literature, e.g.  Sugimoto
\& Fujimoto~(1978), Iben~(1981), Nomoto et al.~(1982a,b), Iben \&
Tutukov~(1991), and more recently Kusakabe et al.~(2011), the
accretion rate in the Chandrasekhar-mass progenitor scenario should be
sufficiently high to avoid a detonation of He.  Recurrent thermal-pulses
during accretion, however, are likely to occur (Iben~1981). This $s$-process rich
material will act as seed for $p$-process nucleosynthesis during the
explosion.  In Figure~10 we present our results obtained from the {\bf
  DDT-a} model ({\it upper panel}) and the {\bf DDT-b} model ({\it
  middle and lower panel}). Both models have been presented in
Section~2, and involve delayed detonations. The $s$-seeds used are the
one presented in the three panels of Figure~3.  The nucleosynthesis
yields for $A \ge$ 40 from model DDT-a (using the $s$-distribution
plotted in Figure~3 and marked there as ST$\times$2) is given in
Table~5.  Since the SN~Ia model used is the same as the one presented
in the previous Section, the resulting nucleosynthesis for nuclei
below the Fe group does not differ significantly.
The production ratio of the synthesized isotopes normalized to the production ratio of 
$^{144}$Sm (the isotope that has the most similar ratio over solar with respect to  $^{56}$Fe over solar) 
are listed in the last column of Table~3 for the 'traditional' 35 $p$-nuclei, and in the last column of 
Table~4 for the other heavy nuclides with important contribution from $p$-process in SN~Ia.
The problem of the production factors of all isotopes listed in Table~4 will be treated separately,
considering the various contributions at the Solar System formation through a Galactic Chemical Evolution treatment as 
in Travaglio et al.~(2004).

In Figure~10 we plot the production factor of each isotope ${\it i}$
normalized to the ratio of $^{56}$Fe produced by the model over 
($^{56}$Fe)$_\odot$. We note that for many of the $p$-isotopes the
overproduction is at the level of $^{56}$Fe. Starting from the
lightest $p$-isotopes, $^{74}$Se, $^{78}$Kr, $^{92,94}$Mo,
$^{96,98}$Ru, $^{102}$Pd, and $^{106,108}$Cd are produced at the level
of $^{56}$Fe (within a factor of 2). In the case of $^{84}$Sr, mainly
$^{85}$Sr($\gamma$,n)$^{84}$Sr is active to produce $^{84}$Sr at $T_{9}$ $\ge$
2.6. We tested the sensitivity of the
production of $^{84}$Sr to the uncertainty of the 
$^{85}$Sr($\gamma$,n)$^{84}$Sr rate. Rauscher \& Thielemann~(2000)
estimate 30\% uncertainty of the Maxwellian Average Cross Section
(MACS) of this rate at $kT$ = 100 keV (the typical temperature for
explosive conditions).  We found that a small change in the cross
section of this reaction changes the final $^{84}$Sr abundance by a
large factor. This has to be carefully taken into account in $p$-process 
nucleosynthesis calculations.. We also get a high production of
$^{86}$Sr from $p$-process nucleosynthesis (almost at the level of
$^{56}$Fe). $^{86}$Sr is a $s$-only isotope, contributed by both
massive stars (weak component) and by low mass AGB stars (main
component). However we find a substantial $p$-process
contributions to $^{86}$Sr.  

Despite the historical problem of reproducing the solar abundances of
$^{92,94}$Mo, $^{96,98}$Ru (these isotopes were neither found 
in comparable abundance with the other $p$-isotopes in SNII
nor in SN~Ia models or in any other stellar site), we find a very
good agreement of the abundances of these isotopes with all the other
$p$-only isotopes, suggesting SNe~Ia as important stellar sites for the
synthesis of heavy and light $p$-nuclei.  For $^{92}$Mo the most
important production channel is $^{93}$Mo($\gamma$,n)$^{92}$Mo.  The
second most important chain is
$^{96}$Ru($\gamma$,$\alpha$)$^{92}$Mo. A small contribution to
$^{92}$Mo comes from the ($p$,$\gamma$) channel. 

$^{94}$Mo is mainly synthesized via the ($\gamma$,n)
photodisintegration chain starting from $^{98}$Mo.  For $T_9 <$ 3
this is almost the only channel to produce $^{94}$Mo, while for $T_9
\ge$ 3 a contribution of $\sim$30\% also comes from
$^{95}$Tc($\gamma$,p)$^{94}$Mo.

Concerning the two Ru $p$-only isotopes, we find that almost 90\% of
$^{96}$Ru is produced in the chain $^{97}$Ru($\gamma$,n)$^{96}$Ru,
with a small contribution from
$^{100}$Pd($\gamma$,$\alpha$)$^{96}$Ru. In contrast, about 50\% of the
$^{98}$Ru is made by $^{99}$Ru($\gamma$,n)$^{98}$Ru, and $\sim$ 50\%
via $^{99}$Rh($\gamma$,p)$^{98}$Ru.

The $p$-contribution to $^{90}$Zr mainly derives from $^{91}$Nb($\gamma$,p)$^{90}$Zr and
$^{91}$Zr($\gamma$,n)$^{90}$Zr. \\  
The other important Zr isotope in this study, $^{96}$Zr, the most sensitive isotope to the neutron
density, comes mainly from the neutron capture channel
$^{95}$Zr(n,$\gamma$)$^{96}$Zr, where the necessary neutrons are
supplied by the $^{22}$Ne($\alpha$,n)$^{25}$Mg reaction.  $^{22}$Ne,
as shown in Figure~5, burns at very low $T_9$ ($\simeq$1.7) in the
outermost layers of the star. Therefore, as we will discuss in more 
detail below, the abundance obtained can be very sensitive to the
modeling of the explosion.

Kusakabe et al.~(2011) followed the $p$-process nucleosynthesis adopting the W7 C-deflagration model by Nomoto et al.~(1984).
When we compare these trends with selected trajectories at $T_\mathrm{9peak}\simeq$3 in Figures 1 and 3 of Kusakabe et al.~(2011) 
our results are quite similar in all respects (see below for more details). We also agree in most nucleosynthetic details 
they discuss. The differences in the total and relative $p$-process yields may be ascribed to the different numerical accuracy and fine 
resolution treatment of the explosive nucleosynthesis in the outermost layers, where carbon-burning occurs and where most of $p$-nuclei are 
synthesized. We fully agree with Kusakabe et al.~(2011) when they wrote that {\it 'the yields are very sensitive to the temperature and density
trajectories'}. The explosive models we follow, especially for the low temperature trajectories, and the technique we employ for the distribution 
and the total number of tracers in the outermost regions are probably a major cause of the relative flat distribution among the light, intermediate 
and heavy $p$-process yields, including the reproduction of the most abundant $p$-nuclei in the Mo and Ru region.
Comparison with previous $p$-calculation presented by Howard et al.~(1991) and with more technical details in Howard \& Meyer~(1993) is somewhat 
hampered by their parameter study and subdivision in only 15 typical trajectories of the outer region, where they based the post-process 
calculations on a SN~Ia delayed detonation models by Khokhlov~(1991).

In order to explain our result, we select two tracers representative for the highest production of $^{92,94}$Mo.
They have been selected with peak temperature $T_\mathrm{9peak}$ = 3.075 and 3.180, respectively, that correspond
(as one can see in Figure~6) to the maximum production of $^{92}$Mo and of $^{94}$Mo. The two tracers are located in two different
zones of the star, one in the outermost zone, with initial $\rho_{peak}\simeq$4.0 $\times$ 10$^7$ g/cm$^{3}$ (Figure~11, upper panel,
dotted line), and the second one a bit more inside the star, with higher initial $\rho_{peak}\simeq$5.5 $\times$ 10$^8$ g/cm$^{3}$ 
(Figure~11, upper panel, solid line). The inner tracer is firstly reached by the deflagration wave, and has and initially 
rather high density. When it is finally reached by the detonation it has much lower densities due to the pre-expansion.
The time on the plot starts at 0.8 sec (the time at 0 sec corresponds to the start of the deflagration wave).
In the other panels of this plot are shown the mass fractions of $p$, n, $^4$He, $^{12}$C, $^{16}$O, $^{20}$Ne, 
$^{23}$Na, $^{22}$Ne, and in the lower panels the mass fraction of $^{92}$Mo and $^{94}$Mo.
Looking at $^{92}$Mo and $^{94}$Mo (lower panels), one can see that, despite of the two different histories of the tracers, 
these isotopes are produced with very similar abundances in these two tracers. Infact the initial density and temperature are 
quite different at t$\simeq$0.0 sec, but, due to the expansion of the star, they are quite similar at the time of the
production of $^{92}$Mo and of $^{94}$Mo. 
The initial $^{12}$C setted with the same value for all tracers, falls abrouptly down when the detonation wave passes through.
Consequently $^{23}$Na is produced, instead $^{20}$Ne at the beginning of C-burning is just mostly dissociated 
producing $\alpha$ and $^{12}$C. 
As to $^{22}$Ne, at the time of $^{92}$Mo and $^{94}$Mo formation, it is strongly reduced with respect to the initial value, with a consequently 
low production of neutrons in the region where the light-$p$ are produced.
After a detailed analysis of the nucleosynthesis that deals the production of the $p$-nuclei, we can state that an accurate treatment of the 
outermost region of the SN~Ia model is at the base of these results. 

Going further to higher mass number, we notice an underproduction in the Cd-In-Sn region. 
The origin of the rare odd isotopes $^{113}$In and $^{115}$Sn (and, related to them,
$^{112}$Sn and $^{114}$Sn) is very debated.  These rare nuclei are
shielded from the two dominant nucleosynthesis processes for heavy
elements, the $s$-process flow proceeds via
$^{112}$Cd$\longrightarrow$$^{113}$Cd$\longrightarrow$$^{114}$Cd$\longrightarrow$$^{115}$In$\longrightarrow$$^{116}$Sn,
thus bypassing the rare nuclei, except for small branchings of the
reaction path at $^{113}$Cd, $^{115}$Cd, and $^{115}$In, which may
contribute to the abundances of $^{113}$In, $^{114}$Sn, and
$^{115}$Sn. The $\beta$-decay chains from the $r$-process region are
shielded by the neutron-rich Cd and In, but may also feed the rare
nuclei via minor decay branchings at $^{113}$Cd and $^{115}$Cd (Rayet,
Prantzos, \& Arnould~1990; Howard, Meyer, \& Woosley~1991; Theis et
al.~1998).  In a previous investigation of the $s$- and $r$-process
components of the Cd, In and Sn isotopes, Nemeth et al.~(1994)
suggested that the rare isotopes in the $A$ = 112 to 115 region may be
produced by a complex interplay of $s$- and $r$-processes.  According
to Dillmann et al.~(2008a) the $s$-process contribution to $^{113}$In
and $^{115}$Sn is excluded as well as the thermally enhanced
$\beta$-decay by an accelerated decay of the quasi-stable $^{113}$Cd
and $^{115}$In during the $s$-process (the mechanism proposed by
Nemeth et al.~1994). The most promising scenario suggested by Dillmann
et al.~(2008a) is related to $\beta$-delayed $r$-process decay
chains. Nevertheless, uncertainties in nuclear physics have to be
taken into account.

Moving to the intermediate $p$-isotopes, we notice $^{138}$La and $^{180m}$Ta far
below the average $p$-nuclei production (by a factor of $\sim$50 and $\sim$8, respectively).  
The astrophysical origin of the two heaviest odd-odd nuclei $^{138}$La and $^{180m}$Ta has been
discussed over the last 30 years (Woosley \& Howard~1978; Beer \&
Ward~1981; Yokoi \& Takahashi~1983; Woosley et al.~1990; Rauscher et
al.~2002), and more recently by Cheoun et al.~(2010). We derive a quite small
contribution from SNe~Ia.  $^{180m}$Ta receives an important
contribution from the $s$-process, due to the branching at $^{179}$Hf, a
stable isotope that becomes unstable at stellar temperatures
(Takahashi \& Yokoi~1987). There is also a second branching at
$^{180m}$Ta due to $^{180}$Hf that at $T_9 \simeq$ 0.3 is not
thermalized and quickly decays to $^{180m}$Ta (Beer \& Ward~1981,
see also Mohr et al.~2007). In core-collapse supernova $^{180m}$Ta is
synthesized by the neutrino-process (Woosley et al.~1990).  Independent
of the production mechanism, according to Mohr et al.~(2007),
freeze-out from thermal equilibrium occurs at $kT\simeq$ 40 keV, and
only $\sim$35\% of the synthesized $^{180m}$Ta survives in the isomeric
state. Consequently in all supernova results, 
the yield obtained so far without accounting of the freeze-out
effect of excited levels discussed by Mohr et al.~(2007) and 
Hayakawa et al.~(2010b) should be decreased by about 2/3. Hayakawa et al. (2010b) positively cited
the Mohr et al.~(2007) evaluation, writing that {\it the isomeric residual population Ratio R}
by  Mohr et al., i.e. R=P$_m$/(P$_g$+P$_m$)= 0.35$\pm$0.4. It was based {\it 'from an estimate of 
the freeze-out temperature without following the time-dependent evolution in detail'}.
The new result by Hayakawa et al.~(2010b), i.e. R=0.38, essentially coincides with R=0.39$\pm$0.1 given 
by Hayakawa et al.~(2010a).
The new problem resides in the conclusion given by Hayakawa et al.~(2010b), {\it 'the main conclusion of the previous 
study by  Hayakawa et al.~(2010a) is thus strenghthned: the solar abundances of  $^{138}$La and  $^{180m}$Ta relative to 
$^{16}$O can be sistematically reproduced by neutrino nucleosynthesis and an electron neutrino temperature of kT close to 4 MeV'}.
Taken at face value, this sentence contrasts with the factc that the $s$-process produces already about 50\% of solar $^{180m}$Ta
(Mohr et al.~2007), whereas no $s$-process $^{138}$La is predicted, and even more with our present $p$-process SN~Ia result 
that further provides extra $p$-contribution to solar $^{180m}$Ta (and to $^{138}$La).
Notice that in Figure~10, there are two $p$-process nuclides at $A$=180, the higher one is $^{180}$W (which in
turn is only produced by the $s$-process at the level of 5\% (see the review by  K\"appeler et al.~2011). 
The conclusion by Hayakawa et al.~(2010b) should however account for the uncertainties in the stellar neutrino cross sections
involved in the production of $^{180m}$Ta and $^{138}$La.

Concerning $^{152}$Gd and $^{164}$Er, we already outlined that both
isotopes are mainly reproduced by the $s$-process (Arlandini et
al.~1999), through the branching at $^{151}$Sm (see updates in Marrone
et al.~2006), and by the branching at $^{163}$Dy, another stable
isotope that becomes unstable at stellar temperatures (Takahashi \&
Yokoi~1987). No production of $^{152}$Gd is expected by the weak s-process 
in massive stars since it is strongly destroyed during
carbon-shell burning (see The et al.~2007 and references therein).
For $^{152}$Gd we find that it is produced by a factor 100 less than the
average $p$-distribution, which confirms the non-$p$ astrophysical
origin of this isotope.  
Also $^{164}$Er is underproduced by one order of magnitude with respect to the
heavy $p$-nuclei. Both $^{152}$Gd and $^{164}$Er are to be classified 
as $s$-only isotopes, not $p$-only.

\subsection{Different $s$-process distributions}

Using the $s$-process seeds plotted Figure~4 (upper and middle panels), corresponding to 
a non flat $s$-distribution, peaked to the lighter $s$-isotopes with $A \sim$ 80$--$90), 
the nucleosynthesis results are shown in the two panels of Figure~12.  
Concerning the $p$-nuclei, we notice that with a $s$-seed distributions flat (see Figure~3, upper 
and lower panel) or peaked to the lighter $s$-isotopes (see Figure~4, upper and middle panel), 
almost all $p$-nuclei scale linearly with metallicity (within a factor of $\sim$2), with the 
exception of the three lightest $p$ nuclei $^{74}$Se, $^{78}$Kr, and $^{84}$Sr, and $^{180m}$Ta, which show a
different behavior. While a decrease of $^{74}$Se, $^{78}$Kr, and
$^{84}$Sr by a factor of $\sim$20 would be expected when changing the
metallicity from solar to $Z$=0.001, we observe a variation by a
factor of $\sim$2 only.  For $^{180m}$Ta decreasing metallicity by a
factor of 20 (from solar down to $Z$=0.001) we obtain an increase of
the $p$-process contribution of a factor of $\sim$10, still far
below the average $p$-nuclei abundance obtained for these models.

A particular attention has to be devoted to the $s$-distribution plotted in Figure~4, lower panel, 
peaked to the heavier $s$-isotopes. For this case, the $p$-nuclei on average are produced at the 
level of $^{56}$Fe, when normalized to solar abundances (see Figure~10, lower panel).

\subsection{$^{208}$Pb as seed of $p$-nuclei}

In order to check in more detail the effect of the $s$-process seed on
the resulting $p$-nuclei, we run a further test for model DDT-a. We
adopt solar $s$-process abundances except for $^{208}$Pb, for which we
use the value from the ST$\times$2 distribution (shown in Figure~3,
upper panel).  If a significant fraction of the seed abundances is
present in form of Bi and Pb isotopes, they are converted to
nuclei of lower mass by photodisintegration sequences starting with
($\gamma$,n) reactions (Dillmann et al.~2008a).  As could be expected,
we obtain an important $p$-contribution (of $\simeq$60\%) to
$^{144}$Sm and $^{196}$Hg using the $^{208}$Pb $s$-enhanced seed. Also
$^{168}$Yb, $^{174}$Hf and $^{190}$Pt get a substantial contribution
(about 25-30\%) this way.  No important contribution derives on the
light $p$-nuclei in this case.

\subsection{Resolution study}

We performed a resolution study for model DDT-a, testing whether the
number of tracer particles used for our calculation is sufficient to
obtain converged nucleosynthesis predictions.  Since we are interested
in the convergence of the $p$-process yields, this could be achieved
in an efficient way (for general convergence tests, see Seitenzahl et
al.~2010). Instead of increasing the number of tracer particles, we
reduced the mass represented by each tracer in the $p$-process region
with the variable tracer mass method (Seitenzahl et al.~2010). The
resolution study consisted of a run of the same DDT-a model, but
computed with variable tracer masses.  The $p$-process mass covered in
the DDT-a standard model is $\sim$0.13 $M_\odot$ with tracer particles
with a constant mass of 2.75 $\times$ 10$^{-5} M_\odot$. In the resolution test
performed we covered $\sim$0.2 $M_\odot$ in the $p$-process
temperature range, and the tracer particles have masses in the range
1.13 $\times$ 10$^{-5} M_\odot$ and 4.90 $\times$ 10$^{-5} M_\odot$. Between the two cases the
resulting $p$-process abundances differ by a few \% only. Therefore,
we can consider our DDT-a model (and all the other models presented in
this paper) well resolved for the $p$-process calculations.

\subsection{Pure deflagration and different strengths of delayed
  detonations: consequences for the $p$-process}

In addition to our `standard case' DDT-a, we performed $p$-process
nucleosynthesis calculations for the other SN~Ia models presented in
Section~2, i.e. pure deflagration, and delayed detonations of
different strenghts. In Figure~13 we show the mass distribution (in
the form of tracer particles) at different $T_{peak}$ (in the
temperature range of $p$-process nucleosynthesis).  In the pure
deflagration case (DEF-a) it is clear from the histogram that a lot of
mass remains unburned (below $T_9 \simeq$ 1) (as was also discussed in
more detail in Travaglio et al.~2004 where only deflagration models
were presented). Consequently, in thisa case we have a factor of
$\sim$10 more $^{22}$Ne which results in a much higher neutron
abundance and, thus, the nuclei most sensitive to the neutron density
(e.g. $^{54}$Fe, $^{58}$Ni, $^{96}$Zr) show a huge overproduction.  It
is also important to note that in the DEF-a model, $^{92}$Mo,
$^{96}$Ru are lower than the DDT-a model (using the same initial
$s$-process distribution) by a factor of $\sim$3, and $^{74}$Se,
$^{78}$Kr and $^{84}$Sr by a factor of $\sim$2. This effect can be
attributed to the different distribution of mass in the two models. In
fact, the amount of mass (corresponding to the number of tracers) at
$T_9 \ge$ 3 in the DDT-a model is a factor of $\simeq$3 higher than in
the DEF-a model.  For $^{168}$Yb, $^{174}$Hf, $^{180}$W, $^{184}$Os
and $^{196}$Hg we observe the opposite effect. We see that in
the DEF-a model these heavy $p$-isotopes are more abundant by a factor
between 2 and 3 with respect to the DDT-a case. From Figure~13 we
notice that the mass in the range $T_9 \sim$2.5 $-$ 2.7 (as shown in
Figure~6 and Figure~7 this is the $T$ zone with the highest production
of these isotopes) is a factor of $\sim$2 lower for the DDT model with
respect to the DEF-a model. From this we conlcude that the mass
distributions and therefore the underlying SN~Ia model can introduce
quite important changes in the final $p$-process nucleosynthesis.

For the pure-deflagration SN~Ia model (DEF-a), the results of the
nucleosynthesis calculations are shown in Figure~14 ({\it lower
  panel}) for all isotopes with $A \ge$40.  The $s$-process initial
seeds used are the ones plotted in Figure~3 (upper panel).  Figure~14
also gives the results for two other DDT models. As was pointed out by
R\"opke \& Niemeyer~(2007), Mazzali et al.~(2007), and Kasen et
al.~(2009), it is possible to cover a wide range of $^{56}$Ni masses
with this class of models.  The variation is introduced by different
ignition geometries that set the strength of the initial deflagration
phase and thus pre-expand the white dwarf prior to detonation
triggering to variable degrees. Apart from the DDT-a model discussed
in previous Sections, we calculated the nucleosynthesis for two
additional delayed detonation models, one with a stronger (DDTs-a) and
one with a weaker (DDTw-a) detonation phase (details of these models
are given in Section~2). For the DDTw-a (Figure~14, upper panel), we
notice a much higher production of the light and the intermediate
$p$-nuclei (up to a factor of $\sim$5 for $^{84}$Sr). This is mainly
connected to the distribution of tracers, for $T_9 \ge$ 3.  In fact,
as shown in Figure~13, with respect to DDT-a the DDTw-a model has 2 to
4 times more tracer particles in the high-T $p$-region, where most of
the light $p$-nuclei are produced. A smaller effect is seen for the
heavier $p$-nuclei. They are produced at lower $T_9$ ($\simeq$2),
where the difference between the two tracer distribution is not as
high as in highest $T_9$ zones. Nevertheless, using for DDT-a and
DDTw-a the same $s$-process seed abundances, we obtain an almost
flat distribution for the resulting $p$-nuclei, including the lightest
ones. It is important to note that in the case DDTw-a we
significantly increase $^{113}$In and $^{115}$Sn, producing them at
the same level as $^{56}$Fe (within a factor of $\sim$2).

For the DDTs-a model, we find generally lower abundances of the
$p$-nuclei. This is due to the fact that much less mass is in the
low-$T$ region where the $p$-process can occur, and most of the
material reaches nuclear statistical equilibrium condition, and mainly
producing $^{56}$Fe (as indicated in Table~2). Nevertheless, on
average the $p$-nuclei show an almost flat distribution (deriving from
the same distribution of the $s$-process seeds).

\section{Conclusions and future work}

We have presented results of detailed nucleosynthesis calculations for
two-dimensional delayed detonation and deflagration SN~Ia models,
focusing in particular on $p$-process nucleosynthesis.  We used
initial abundances of $s$-nuclei synthesized during the past
AGB-history of the WD and during the mass accretion phase. During the
late AGB phase of the companion, about 0.1 $M_\odot$ of the CO-core
become enriched in $s$-process elements by the effect of recurrent
He-shell thermal instabilities. About 1000 years before the explosion,
the simmering phase induced by central C-burning dilutes the $s$-rich
material by a factor of $\sim$ 10 over the whole WD. The corresponding
$p$-process yield is negligible when integrated over the whole
ejecta. In contrast, the $s$-seeds synthesized during the the WD mass
accretion phase may give rise to an average $p$-process overabundance
in the ejecta comparable with that of $^{56}$Fe. Neutrons fluxes
available for the $s$-process in the accreted material rely on the
activation of the $^{13}$C($\alpha$,n)$^{16}$O reaction during the
convective He flashes. This applies under the assumption that a small
amount of protons are ingested in the top layers of the He intershell,
as was suggested by Iben~(1981).  Protons are captured by the
abundant $^{12}$C and converted into $^{13}$C via
$^{12}$C($p$,$\gamma$)$^{13}$N($\beta^+$$\nu$)$^{13}$C at $T$ $\sim$ 1
$\times$10$^8$ K. Neutrons are then released in the bottom region of
the convective He intershell, by
$^{13}$C($\alpha$,n)$^{16}$O. Differently from the previous AGB phase,
where both the formation of the $^{13}$C-pocket and the subsequent
release of neutrons occur radiatively in the interpulse phase, here
the $s$-process is made directly in the convective shell during a
thermal instability, similarly to the {\it plume mixing} by Ulrich~(1973). 
The results of our post-process
calculations should be considered as preliminar, given the general
difficulty of following with full evolutionary codes the peculiar
conditions of thermal-pulses during mass accretion phase.

In the cases of $s$-seeds from the mass accretion phase, we analyzed
different SN~Ia models, i.e. delayed detonations of different strength,
and a pure-deflagration model.  For all these models we explored
different $s$-process distributions, and their consequences for the
$p$-process.  We also investigated a metallicity effect of the
$p$-process nucleosynthesis in this scenario, considering models with
solar metallicity and 1/20 solar.  Despite the fact that studies of
$s$-process nucleosynthesis during the accreting WD phase are lead by
the need to clarify the effective $s$-process distribution in these
particular conditions, the results presented in this paper for
$p$-process nucleosynthesis are quite significant.  Note that a flat
$s$-seed distribution directly translates into a flat $p$-process
distribution whose average production factor scales linearly with the
adopted level of the $s$-seeds.  In contrast to previous work on
$p$-process nucleosynthesis in SN~Ia (e.g. Kusakabe et al.~2005, 2011;
Goriely et al.~2002, 2005), we demonstrated that we can produce almost
all the $p$-nuclei with similar enhancement factors relative to
$^{56}$Fe, including the puzzling light $p$-nuclei $^{92,94}$Mo,
$^{96,98}$Ru.  We found that only the isotopes $^{113}$In, $^{115}$Sn,
$^{138}$La, $^{152}$Gd, and $^{180m}$Ta to diverge from the average
$p$-process production.  Among them, $^{152}$Gd, and $^{180m}$Ta have
an important contribution from $s$-process in AGB stars (Arlandini et
al.~1999) or the neutrino process in SNII (Woosley et al.~1990; Wanajo 
et al.~2011a). Both $^{113}$In and $^{115}$Sn are not fed by the $p$-process 
nor by the $s$-process. For them we refer to the discussion by Dillmann 
et al.~(2008b).

As far as the Galactic chemical evolution of $p$-nuclei is concerned
our results lead to the following very preliminar conclusions. Given
the assumption of different $s$-seed distributions (see Figs.~3 and ~4),
for both solar metallicity and 1/20 than solar, we could show that the 
$p$-nuclei on average are produced at the level of $^{56}$Fe, when 
normalized to solar abundances (as shown in Figure~10). 
Even taking a fixed choice of the $^{13}$C-pocket at different metallicities,
since the $s$-seed distributions peaked at heavier mass number
are in average dominant, we may infer that the ($p$/$^{56}$Fe)/($p$/$^{56}$Fe))$_\odot$
ratio is always constant. This suggest the primary nature of $p$-process.
This aspect will be examined thoroughly in a forthcoming paper.
From the hyphothesis that SNe~Ia are responsible for 2/3 of the solar
$^{56}$Fe, and by assuming that our DDT-a model represents the typical
SN~Ia with a frequency of $\sim$70\% of all SNe~Ia (Li et al.~2011), we
conclude that they can be responsible for about 50\% of the solar
abundances of all $p$-nuclei. Instead, if we consider an average between 
DDTw-a and DDTs-a models, considering that they represent $\sim$10\% of all 
SNe~Ia (Li et al.~2011), still with a flat $s$-seed distribution (see Figure~14),
we obtain that they can account for 75\% of the solar abundances of all $p$-nuclei.
Of course, these are first rough estimates of how SNe~Ia, in principle,
can contribute to the Solar System composition of $p$-nuclei. 
These results must also take into account that Type II SNe can be potentially 
important contributors to the galactic $p$-nuclei. Rayet et al.~(1995) shown that
about 1/4 of the Solar System $p$-nuclei can be attributed to SNII explosions.
Later on, Woosley \& Heger~(2007) showed that $p$-nuclei can have an appreciable 
contribution at the solar composition from explosive neon and oxygen burning for 
the mass number greater than 130, but are underproduced for lighter masses. 
As recalled in Section~1, recent works on ${\nu}p$-process in SNII show that maybe 
a non negligible contribution can come from this process to the light $p$-isotopes.

A more thorough analysis of the role of SNe~Ia in the solar composition of $p$-nuclei 
is planned.

Finally we note that recent works (e.g. Sim et al.~2010; Fink et
al.~2010, and references therein) discussed the fact that the
explosion of sub-Chandrasekhar mass white dwarfs via the double
detonation scenario is a potential explanation for Type Ia supernovae
(but see Woosley \& Kasen~2010). Again, the possibility of $p$-process
production will be explored in a forthcoming paper.

All the Tables with the yields obtained for different models presented
in this work are available on request.

 \acknowledgments 
We thank S. Cristallo and O. Straniero for many discussions on the feasibility of
the $s$-process during mass accretion. We thank U. Battino for testing the
production channels of the various $p$-isotopes in the occasion of his
Master Degree. We thank P. Mohr for an extreme detailed update
of the nucleosynthesis of  $^{180m}$Ta and $^{138}$La in explosive SN conditions.
A deeply thank to the anonymous referee for his comments that helped much to improve the paper.
This work has been supported by B2FH Association. The numerical calculations have been also 
supported by Regione Lombardia and CILEA Consortium through a LISA Initiative (Laboratory for 
Interdisciplinary Advanced Simulation) 2010 grant. This work was partially supported by the Deutsche
 Forschungsgemeinschaft via the Transregional Collaborative Research
 Center ``The Dark Universe'' (TRR~33), the Emmy Noether Program (RO
 3676/1-1), and the Excellence Cluster ``Origin and Structure of the
 Universe'' (EXC~153).

\clearpage

\begin{figure}
\centerline{\includegraphics[width=\textwidth]{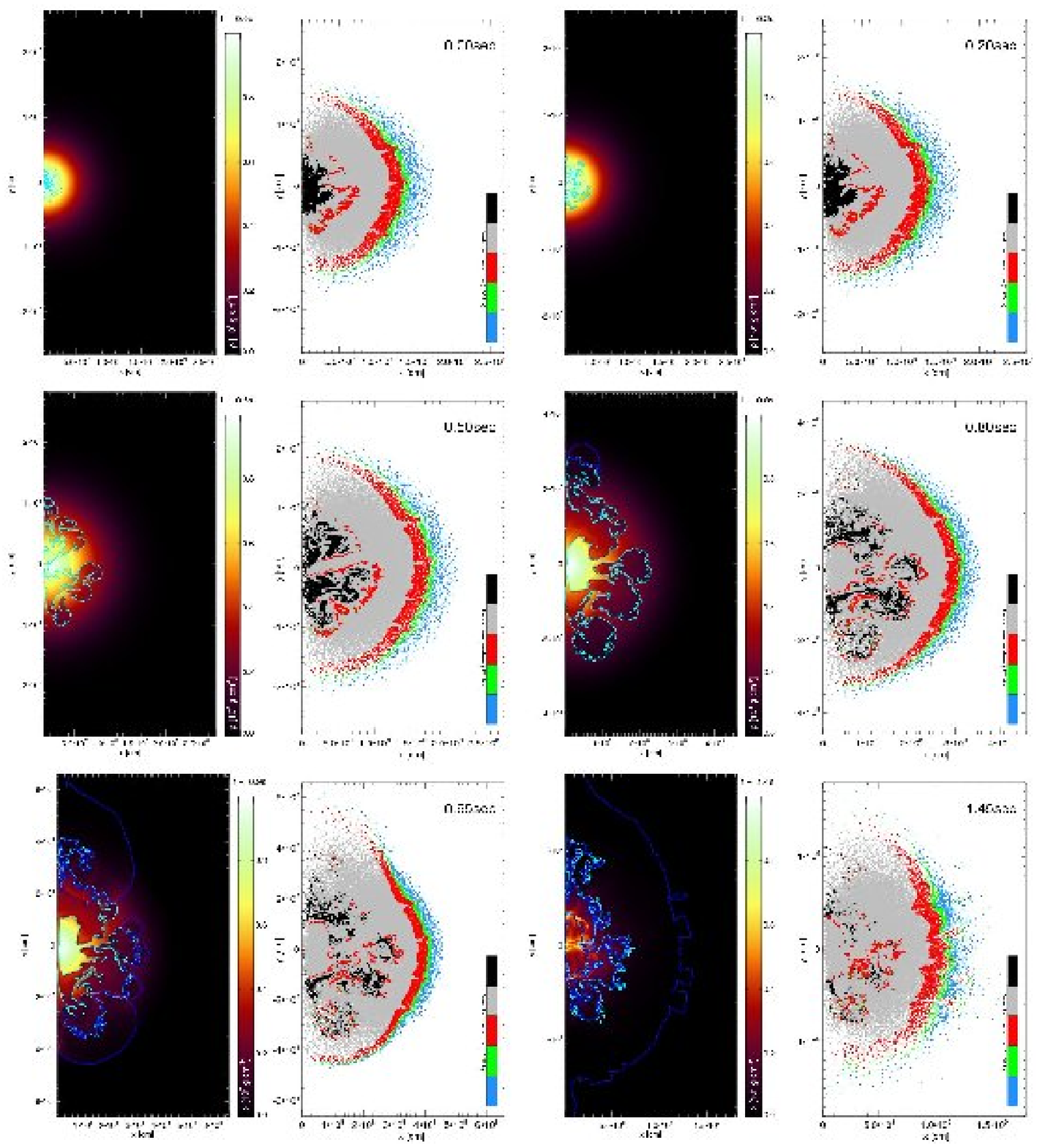}}
\caption{{\small Snapshots from model DDT-a at $0.0\,\mathrm{s}$, $0.2\,\mathrm{s}$, $0.5\,\mathrm{s}$, $0.8\,\mathrm{s}$, $0.95\,\mathrm{s}$ and
$1.45\,\mathrm{s}$ after ignition. On the left, the hydrodynamic evolution is illustrated by color-coded density and the locations of
the deflagration flame (cyan contour) and the detonation front (blue contour). In the model, the first detonation triggers at
$0.755\,\mathrm{s}$ after ignition. The plots on the right hand side show the tracer distribution. While the locations correspond to the
current time, the color coding is according to the maximum temperature reached during the entire explosion: {\it Black} tracers
peak with $T_9^\mathrm{peak} >$ 7.0; {\it grey} tracers with $3.7 < T_9^\mathrm{peak} < 7.0$; tracers marked 
in {\it blue} ($1.5 < T_9^\mathrm{peak} < 2.4$), {\it green} ($2.4 < T_9^\mathrm{peak} < 3.0$) and {\it red} ($3.0 < T_9^\mathrm{peak} < 3.7$) are reach 
peak temperatures in ranges where $p$-process nucleosynthesis is possible.}}\label{fig1}
\end{figure}

\begin{figure}
\includegraphics[scale=.60]{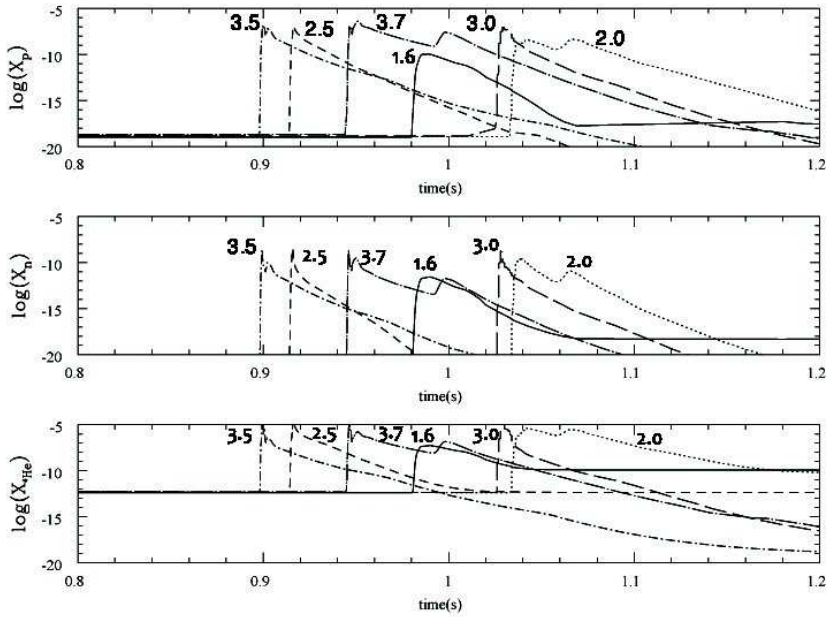}
\caption{Time evolution of mass fractions of proton ({\it upper panel}), neutron ({\it middle panel}), and $^{4}$He ({\it lower panel}), for selected tracers at different 
peak temperatures (indicated with labels in the panels), i.e. $T_9$ = 1.6, 2.0, 2.5, 3.0, 3.5 and 3.7.}\label{fig2}
\end{figure}

\begin{figure}
\includegraphics[angle=-90,scale=.60]{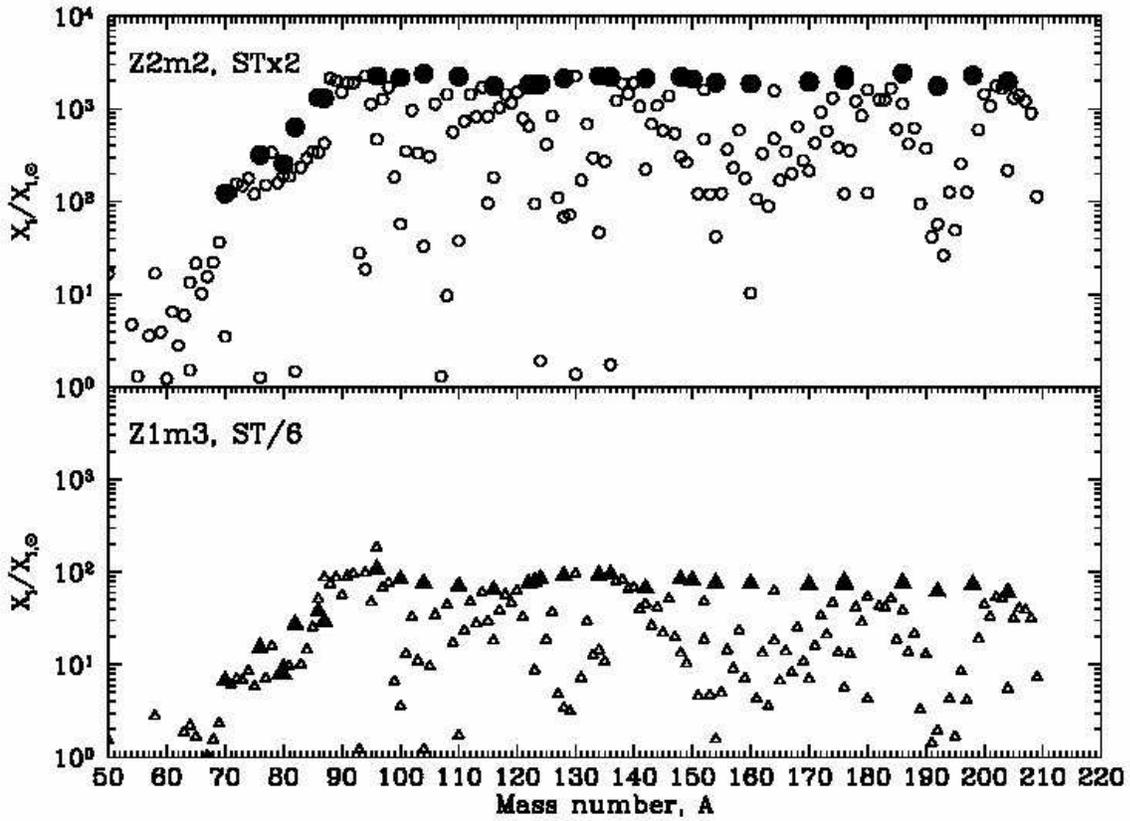}
\caption{Distribution of initial seed abundances relative to solar for $Z$=0.02, STx2 case (see text), in the {\it upper panel}, and for $Z$=0.001, ST/6 case (see text), in the 
{\it lower panel}. Filled dots and triangles are for $s$-only isotopes.}\label{fig3}
\end{figure}

\begin{figure}
\includegraphics[angle=-90,scale=.60]{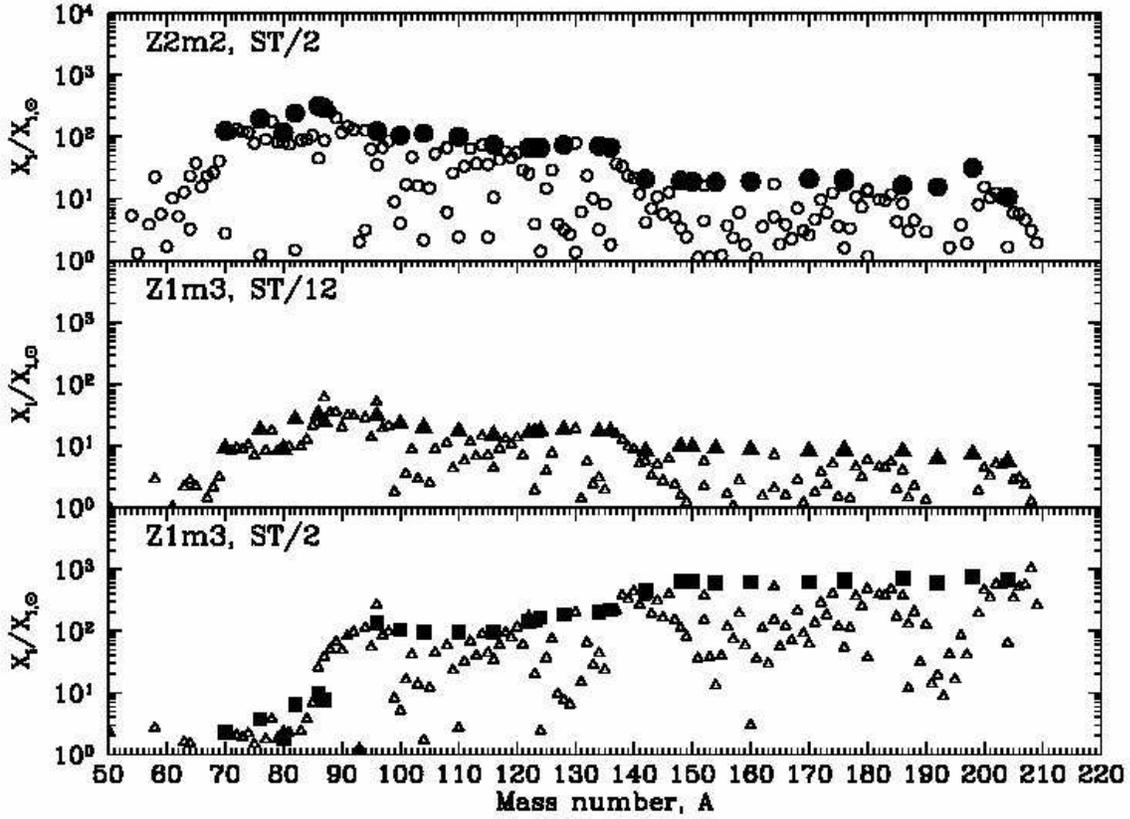}
\caption{Distribution of initial seed abundances relative to solar for $Z$=0.02, ST/2 case (see text), in the {\it upper panel}, for $Z$=0.001, ST/12 case (see text), in 
the {\it middle panel}, and for $Z$=0.001, ST/2 case (see text), in the {\it lower panel}. 
Filled dots, triangles, and squares  are for $s$-only isotopes.}\label{fig4}
\end{figure}

\begin{figure}
\includegraphics[angle=-90,scale=.60]{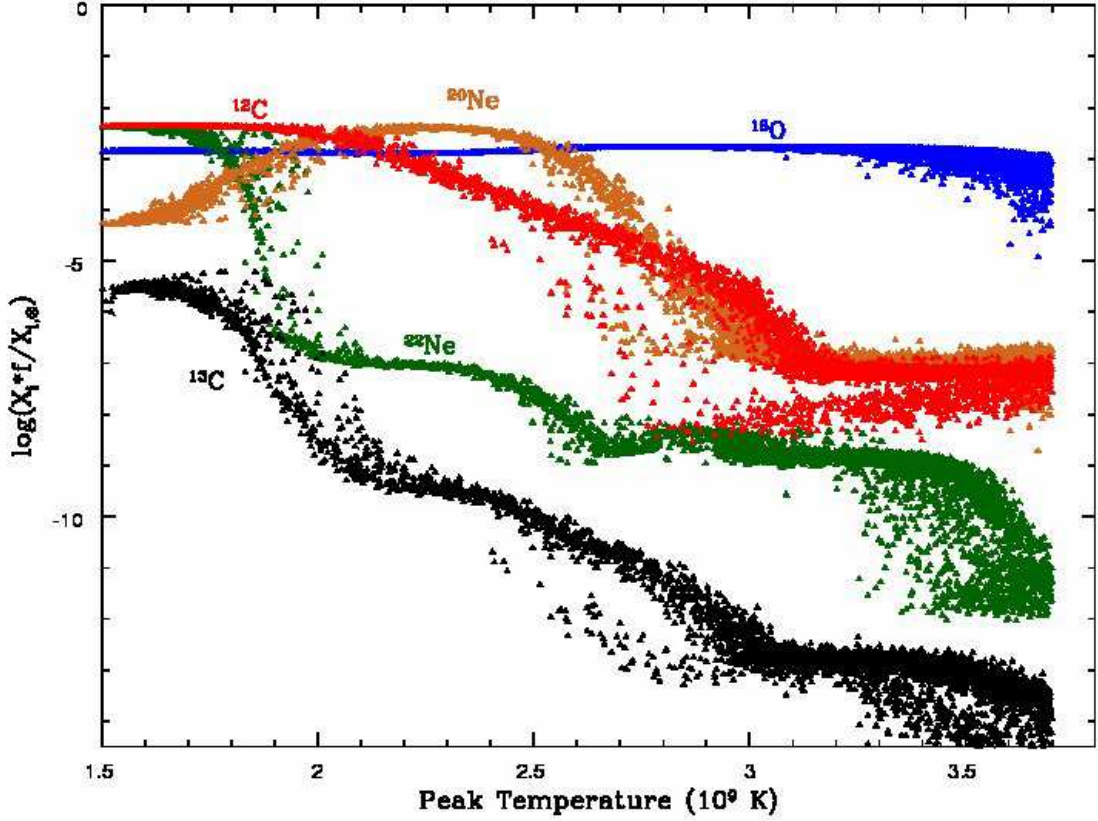}
\caption{Abundance of $^{12}$C ({\it red}), $^{16}$O ({\it blue}), $^{22}$Ne ({\it green}), $^{20}$Ne ({\it cyan})
for tracers selected in the peak $T$ range that allowed $p$-process nucleosynthesis. This is shown for {\bf DDT-a}.
All the abundances $X_i$ of the three panels are for each tracer, and the $f$ factor in the plot 
is for $M_\mathrm{WD}$(= 1.407 M$_\odot$)/N$_{tracers}$ (=51200), i.e. the mass of each tracer.}\label{fig5}
\end{figure}

\begin{figure}
\includegraphics[angle=-90,scale=.60]{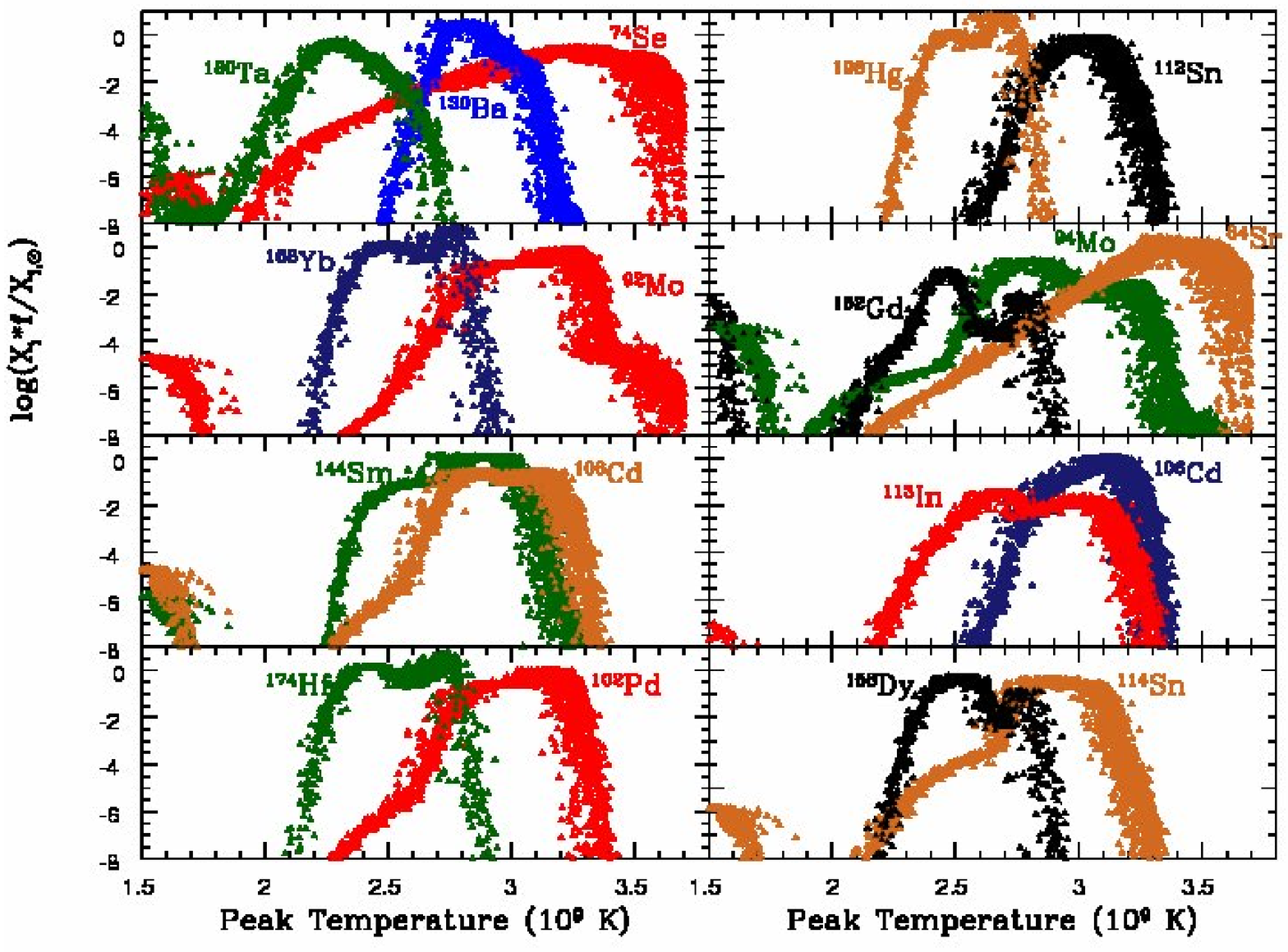}
\caption{Abundances vs. peak $T$, for tracers selected in the 
$T$ range that allowed $p$-process nucleosynthesis. This is shown for {\bf DDT-a}. The $f$ factor is the same of Figure~5.}\label{fig6}
\end{figure}

\begin{figure}
\includegraphics[angle=-90,scale=.60]{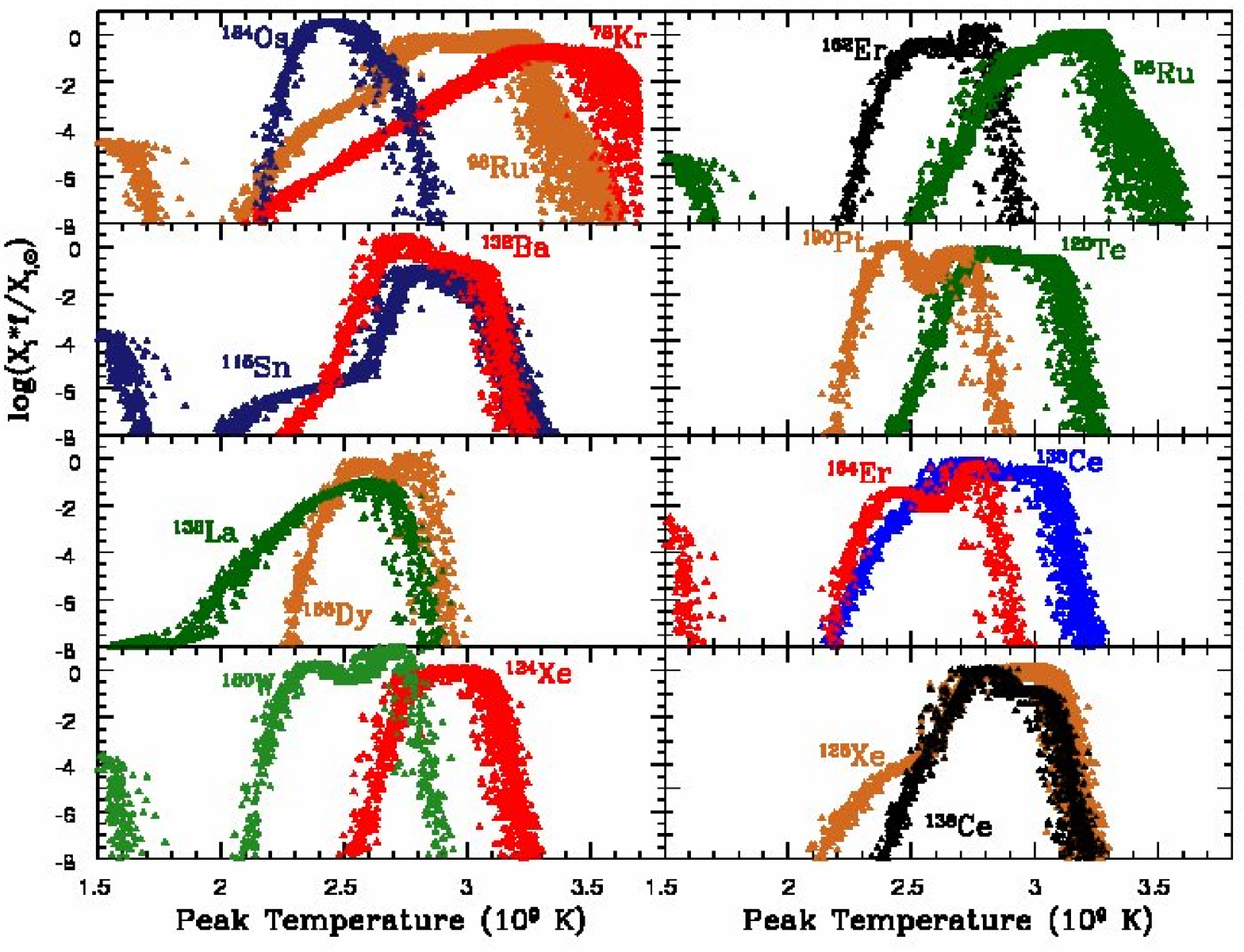}
\caption{Abundances vs. peak $T$, for tracers selected in the 
$T$ range that allowed $p$-process nucleosynthesis. This is shown for {\bf DDT-a}.The $f$ factor is the same of Figure~5.}\label{fig7}
\end{figure}

\begin{figure}
\includegraphics[angle=-90,scale=.60]{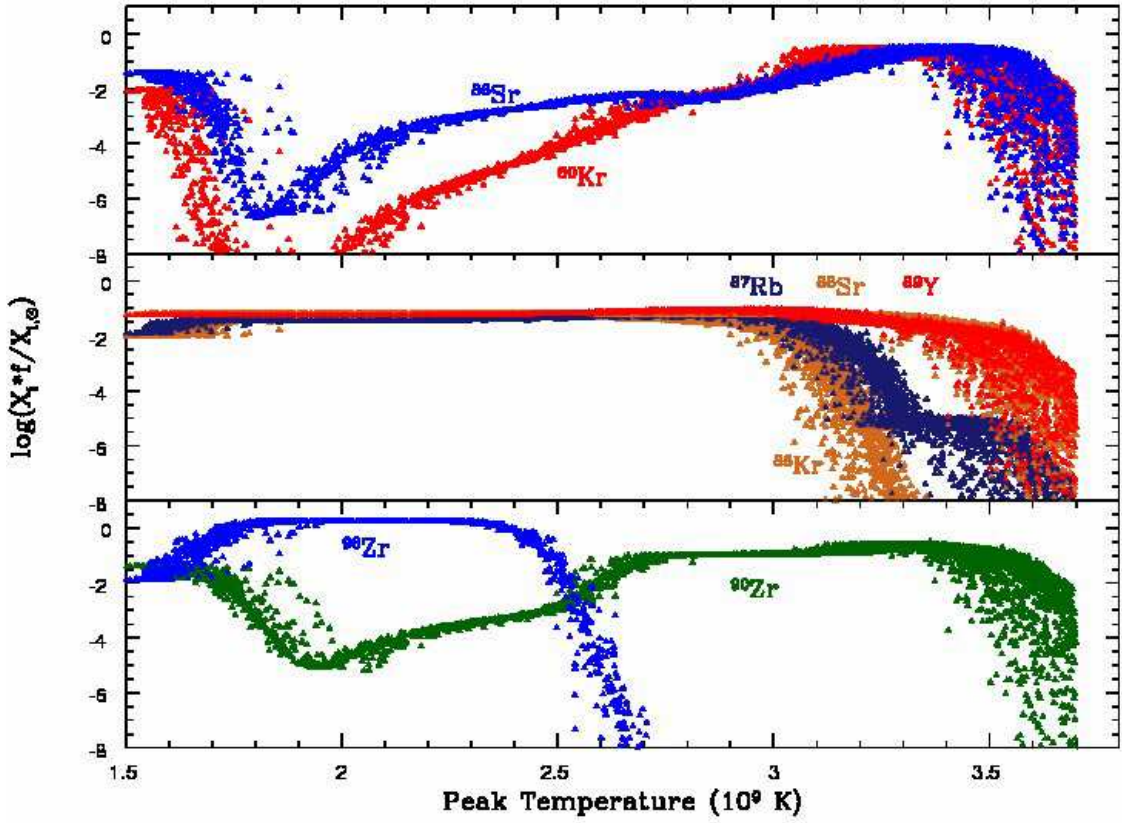}
\caption{Abundances vs. peak $T$, for tracers selected in the 
$T$ range that allowed $p$-process nucleosynthesis. This is shown for {\bf DDT-a}. The selection of the isotopes in this Figure corresponds
of the isotopes listed in Table~2. The $f$ factor is the same of Figure~5.}\label{fig8}
\end{figure}

\begin{figure}
\includegraphics[angle=-90,scale=.60]{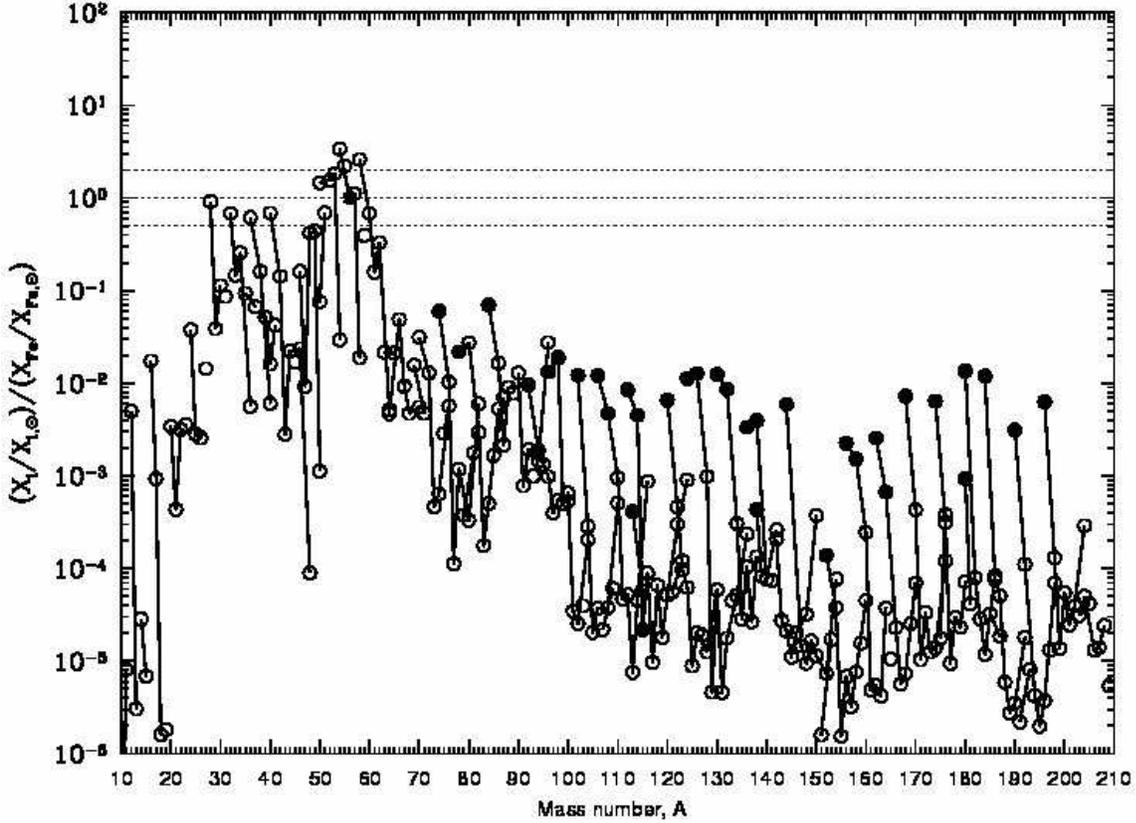}
\caption{Nucleosynthesis yields (production factors normalized to Fe) obtained using 51200 
tracer particles in the 2D {\bf DDT-a} model (as described in the text). The $^{56}$Fe mass fraction obtained is 0.584 M$_\odot$. 
The $s$-process enrichment for this test is considered due only to the AGB phase progenitor of the WD. 
{\it Filled dots} are for the $p$-only isotopes (as defined in Table~1. {\it Diamond} is for $^{56}$Fe.}\label{fig9}
\end{figure}

\begin{figure}
\includegraphics[angle=-90,scale=.60]{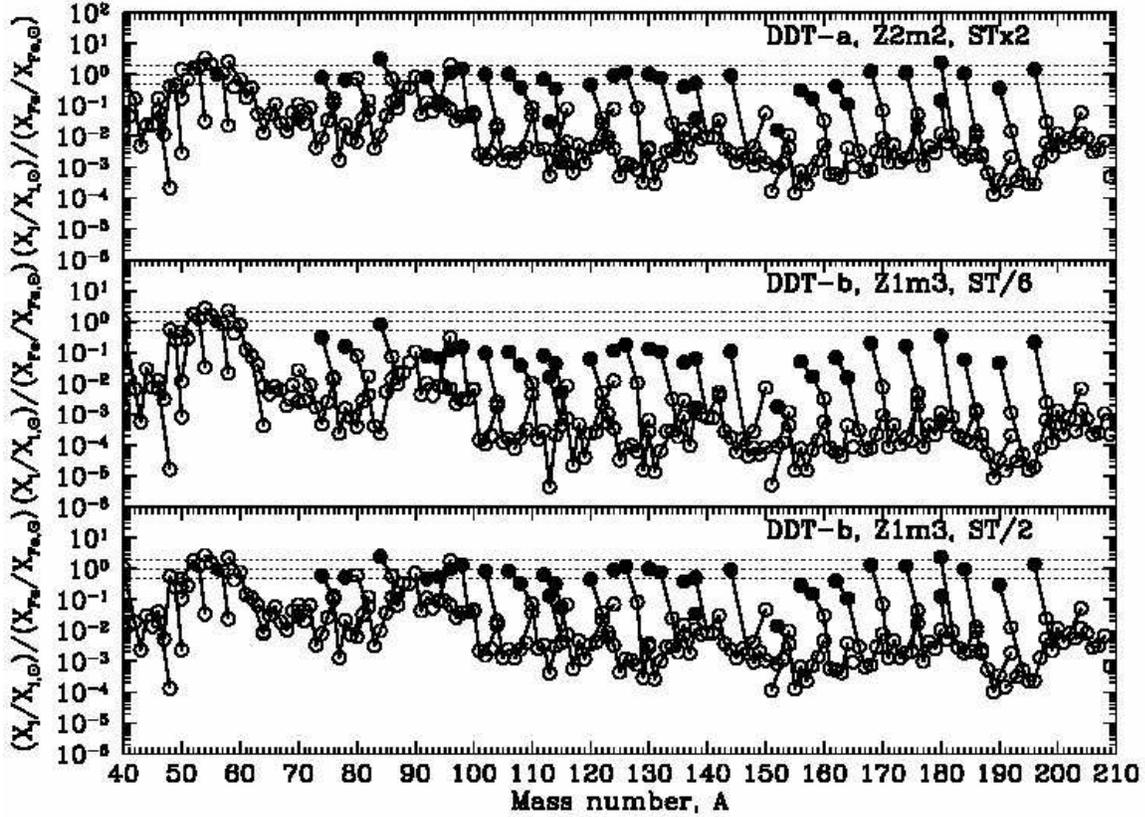}
\caption{Nucleosynthesis yields (production factors normalized to Fe) obtained using 51200 
tracer particles in the 2D {\bf DDT-a} model (as described in the text). The $s$-process enrichment for this case has been considered in the accreted mass with solar 
metallicity ({\it upper panel}) and 1/20 than solar ({\it middle and lower panel}) {\bf DDT-b}, with the $s$-seed distribution plotted in Figure~3 (upper and lower 
panel) and in Figure~4 (lower panel). See text for explanation of the the $^{13}$C-pocket strengths adopted.}\label{fig10}
\end{figure}

\begin{figure}
\includegraphics[scale=.60]{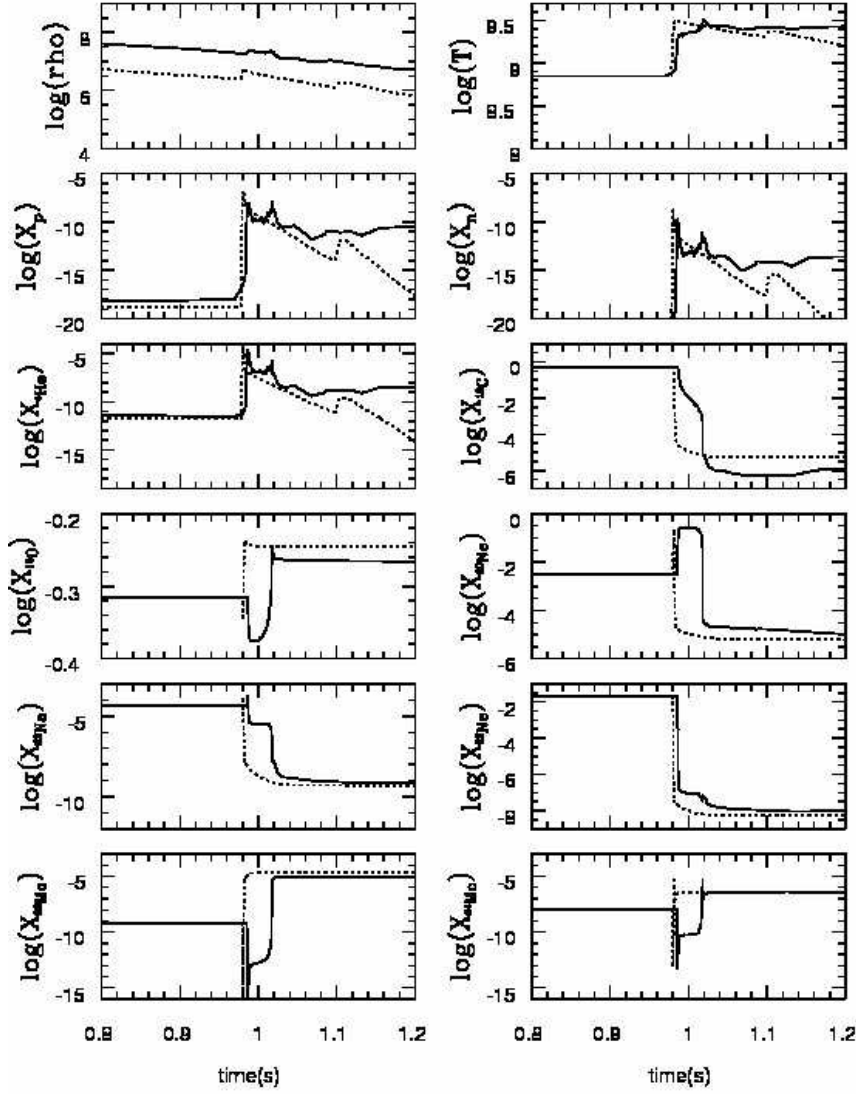}
\caption{Time evolution of density and temperature (upper panels), of mass fractions of proton and neutron ({\it second panels from the top}), 
$^{4}$He and $^{12}$C ({\it third panels from the top}), $^{16}$O and $^{20}$Ne ({\it fourth panels from the top}), $^{23}$Na and $^{22}$Ne ({\it 
second panels from the bottom}), and $^{92}$Mo and $^{94}$Mo ({\it lower panels}). The {\it solid} and {\it dotted} lines are for two different 
tracers.}\label{fig11}
\end{figure}

\begin{figure}
\includegraphics[angle=-90,scale=.60]{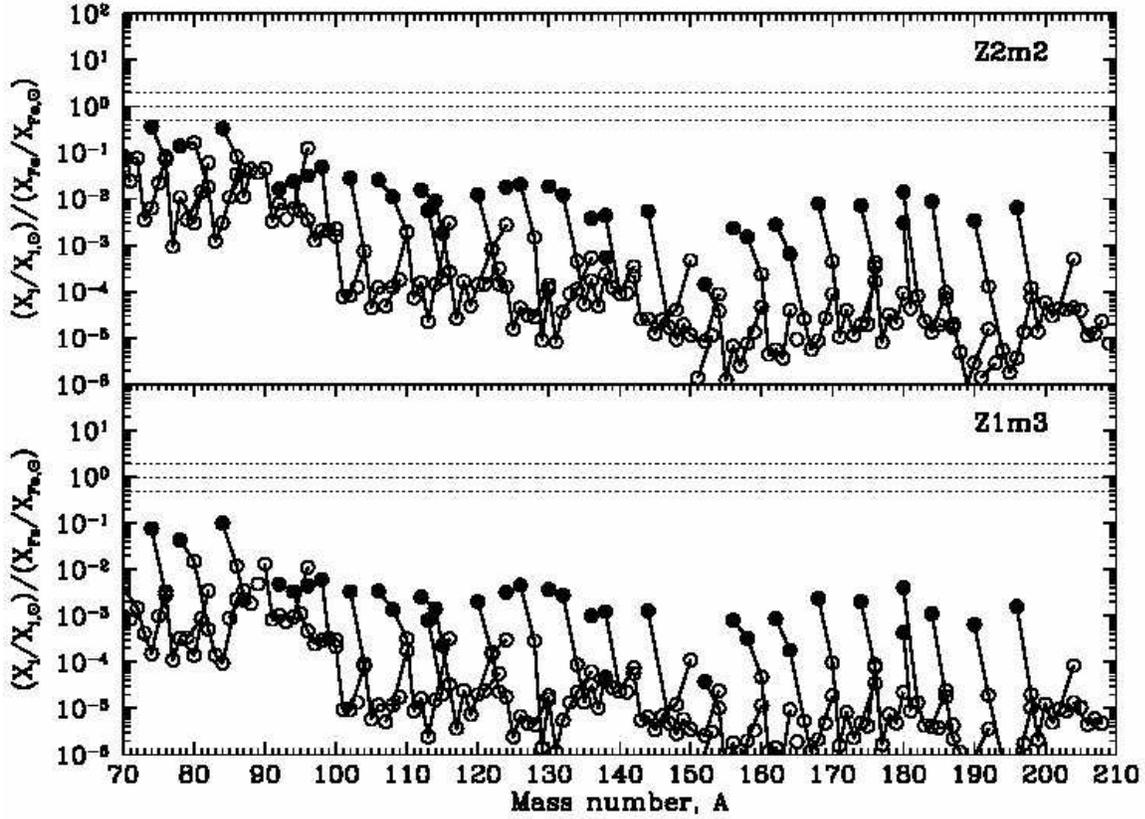}
\caption{Nucleosynthetic yields (production factors normalized to Fe) obtained using 51200 
tracer particles in the 2D {\bf DDT-a} model (as described in the text). The $s$-process enrichment in the accreted mass for this case has been considered with solar 
metallicity ({\it upper panel}) and 1/20 than solar ({\it lower panel}) {\bf DDT-b}, with the non-flat $s$-seed distribution plotted in Figure~4 (upper and middle 
panel, respectively).}\label{fig12}
\end{figure}

\begin{figure}
\plotone{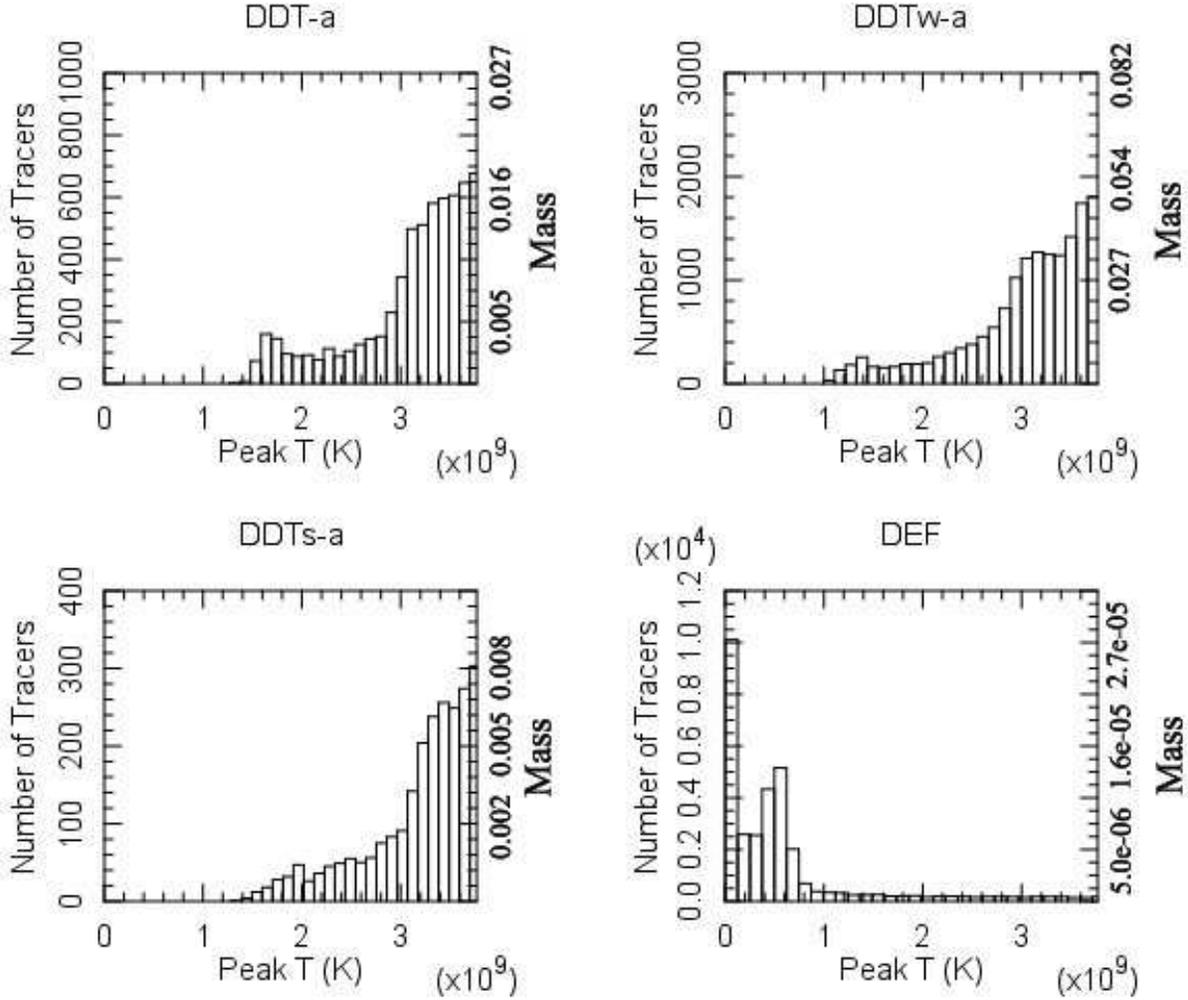}
\caption{Mass distribution (plotted in the form of number of tracers) as a function of peak temperature. The {\it upper left} panel is for the DDT standard model, the 
{\it upper right} panel is for the weaker DDT model, the {\it lower left} panel is for the stronger DDT model, and the {\it lower right} panel is for the deflagration 
model. On the right side of each box is reported the mass distribution (in $M_\odot$) as a function of peak temperature.}\label{fig13}
\end{figure}

\begin{figure}
\includegraphics[angle=-90,scale=.60]{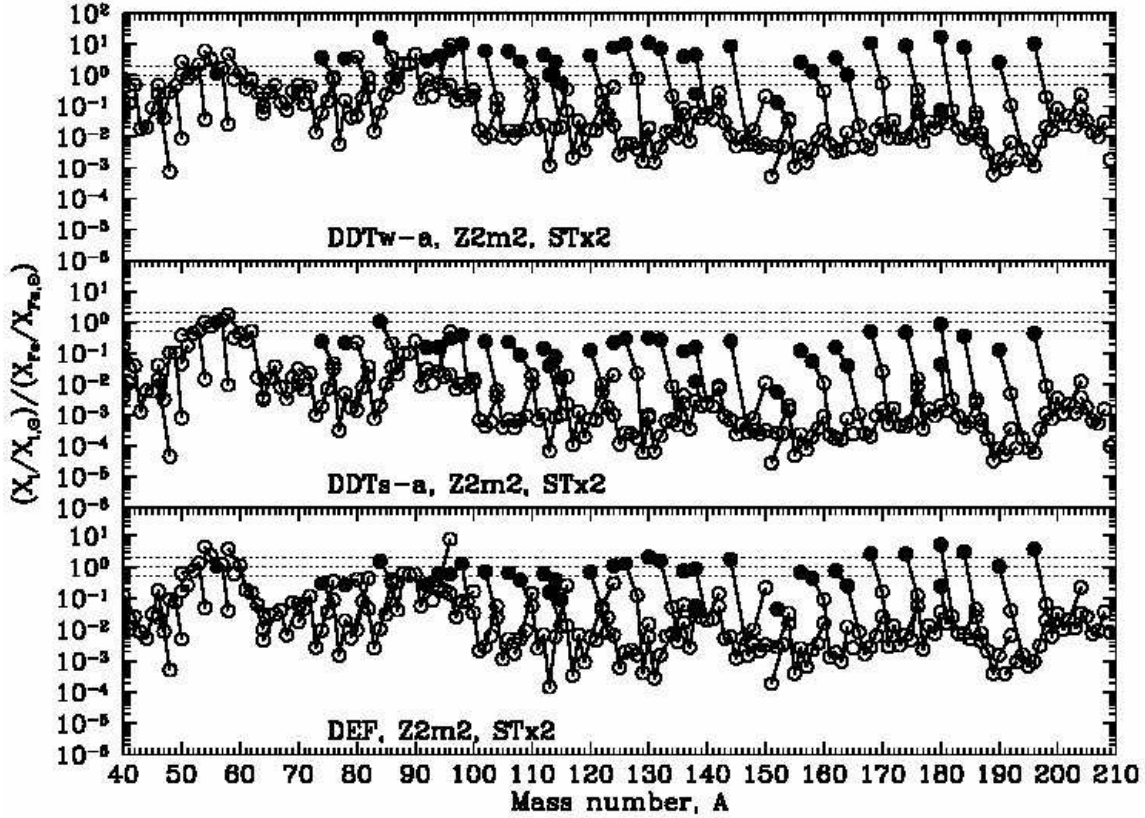}
\caption{Nucleosynthesis yields (production factors normalized to Fe) obtained using 51200
tracer particles in the 2D weaker ({\it upper panel}, DDT-w),  and stronger ({\it middle panel}, DDT-s) delayed detonation models, and pure deflagration model ({\it 
lower panel}, DEF-a) (as described in the text). The $s$-process enrichment for this case is considered in the accreted mass, with the
distribution plotted in Figure~6, upper panel.}\label{fig14}
\end{figure}

\clearpage

\begin{deluxetable}{lccccc}
\tabletypesize{\scriptsize}
\tablecaption{Results of the hydrodynamic SN~Ia explosion simulations: asymptotic kinetic energy $E_\mathrm{kin}^{\mathrm{asym}}$ and final composition (iron group elements: IGE, intermediate-mass elements: IME); all masses are given in solar masses $M_\odot$ \label{table_hydro}}
\tablewidth{0pt}
\tablehead{
\colhead{model} & $E_\mathrm{kin}^{\mathrm{asym}}$ [$10^{51} \, \mathrm{erg}$] & $M(\mathrm{IGE})$ & $M(\mathrm{IME})$ & $M(\mathrm{C})$ & $M(\mathrm{O})$}
\startdata
DEF-a  & $0.3522$ & $0.497$ & $0.081$ & $0.357$ & $0.472$ \\
DDT-a  & $1.3027$ & $0.767$ & $0.509$ & $0.018$ & $0.113$ \\
DDT-b  & $1.2903$ & $0.759$ & $0.522$ & $0.018$ & $0.114$ \\
DDTw-a  & $1.1142$ & $0.531$ & $0.644$ & $0.046$ & $0.187$ \\
DDTs-a  & $1.5017$ & $1.142$ & $0.215$ & $0.007$ & $0.044$ \\
\enddata                                                                 
\end{deluxetable}

\begin{deluxetable}{ccccccc}
\tabletypesize{\scriptsize}
\tablecaption{SN~Ia models: main species}
\tablewidth{0pt}
\tablehead{
\colhead{Species} & \colhead{DDT-a} & \colhead{DDT-b} &  \colhead{DEF-a} & \colhead{DDTw-a} & \colhead{DDTs-a} & \colhead{W7$^a$}}
\startdata
$^{56}$Fe  & 0.583 & 0.487 & 0.317 & 0.301 & 0.986 & 0.669 \\
$^{12}$C  & 0.016 & 0.017 & 0.354 & 0.035 & 0.002 & 0.050 \\
$^{16}$O  & 0.144 & 0.152 & 0.538 & 0.39 & 0.030 & 0.140 \\
$\bigtriangleup M_p^{b}$ & 0.130 & 0.114 & 0.090 & 0.380 & 0.06 & \\
\enddata                                                                 

\vspace{0.1em}
\hspace{9em}$^{(\rm a)}$ -- Iwamoto et al.~(1999)
\vspace{0.1em}

\hspace{9em}$^{(\rm b)}$ -- Fraction of the mass of the star (in $M_\odot$) where $p$-process nucleosynthesis occurs for the SN~Ia models presented in this work.
\vspace{0.1em}

\end{deluxetable}

\begin{deluxetable}{cccc}
\tabletypesize{\scriptsize}
\tablecaption{$p$-nuclides} 
\tablewidth{0pt}
\tablehead{
\colhead{Isotope} & \colhead{\% Isotopic abundance} &  \colhead{Note} &  \colhead{(X$_i$/X$_{i,\odot}$)/($^{144}$Sm/$^{144}$Sm$_\odot$) $^{\rm (a)}$}} 
\startdata
$^{74}$Se  & 0.87 &  & 0.846 \\
$^{78}$Kr  &  0.354 &  & 0.716 \\
$^{84}$Sr  &  0.56 &  & 3.463 \\
$^{92}$Mo  &  15.84 &  & 0.880 \\
$^{94}$Mo  &  9.04 &  & 0.149 \\
$^{96}$Ru  &  5.51 &  &  1.244 \\
$^{98}$Ru  &  1.87 &  & 1.643 \\
$^{102}$Pd &  0.96 &  & 1.092 \\
$^{106}$Cd &  1.215 &  & 1.129 \\
$^{108}$Cd &  0.875 &  & 0.410 \\
$^{113}$In &  4.28 & 1 & 0.032 \\
$^{112}$Sn &  0.96 &  & 0.751 \\
$^{114}$Sn &  0.66 &   & 0.372 \\
$^{115}$Sn &  0.35 & 1 & 0.001 \\
$^{120}$Te &  0.089 &  & 0.506 \\
$^{124}$Xe &  0.126 &  & 0.997 \\
$^{126}$Xe &  0.115 &  & 1.289 \\
$^{130}$Ba &  0.101 &  & 1.118 \\
$^{132}$Ba &  0.0097 &  & 0.838 \\
$^{138}$La &  0.091 &   & 0.041 \\
$^{136}$Ce  &  0.193 &  & 0.425 \\
$^{138}$Ce  &  0.25 &  & 0.551 \\
$^{144}$Sm  &  3.09 &  & 1.000 \\
$^{152}$Gd  &  0.20 & 2 & 0.017  \\
$^{156}$Dy  &  0.0524 &  & 0.332  \\
$^{158}$Dy &  0.0902 &  & 0.178  \\
$^{162}$Er &  0.136 &  & 0.433  \\
$^{164}$Er &  1.56 & 3  & 0.116  \\
$^{168}$Yb &  0.135 &  & 1.353  \\
$^{174}$Hf &  0.18 &  & 1.230  \\
$^{180m}$Ta &  0.0123 & 4 & 0.156  \\
$^{180}$W  &  0.135 &  & 2.633  \\
$^{184}$Os  &  0.018 &  & 1.156  \\
$^{190}$Pt &  0.0127 &  & 0.385  \\
$^{196}$Hg &  0.146 &  & 1.581  \\

\enddata      

\hspace{9em}$^{(\rm 1)}$ -- $^{113}$In and $^{115}$Sn, see text for discussion. Indication for $r$-process contribution (Dillmann et al.~2008) are discussed in the text.
\vspace{0.1em}
                                                           
\hspace{9em}$^{(\rm 2)}$ -- $^{152}$Gd, $s$-only isotope (Arlandini et al.~1999), due to the radiogenic $s$-branching from $^{151}$Sm.
\vspace{0.1em}

\hspace{9em}$^{(\rm 3)}$ -- $^{164}$Er, $s$-only isotope (Arlandini et al.~1999), due to the branching at $^{163}$Dy that is stable at terrestrial
conditions, but becomes unstable at stellar temperatures (Takahashi \& Yokoi~1987).
\vspace{0.1em}

\hspace{9em}$^{(\rm 4)}$ -- $^{180m}$Ta, $s$-only isotope (Arlandini et al.~1999) due to the branching at $^{179}$Hf, is stable at terrestrial conditions and becomes 
unstable at stellar temperatures (Takahashi \& Yokoi~1987). The $s$-process feeds $\sim$ 50\% of the solar $^{180m}$Ta. Another substantial contribution to this isotope comes from 
$\nu$$p$-process in SNII (Woosley et al.~1990).
\vspace{0.1em}

\vspace{0.1em}
\hspace{9em}$^{(\rm a)}$ -- Synthesized mass in DDT-a model normalized to $^{144}$Sm. For the results of this column, see Section~6.2.
\vspace{0.1em}

\end{deluxetable}

\begin{deluxetable}{cccc}
\tabletypesize{\scriptsize}
\tablecaption{$s$-nuclides with $p$-contribution} 
\tablewidth{0pt}
\tablehead{
\colhead{Isotope} & \colhead{\% Isotopic abundance $^{\rm (a)}$} &  \colhead{Note} &  \colhead{ (X$_i$/X$_{i,\odot}$)/($^{144}$Sm/$^{144}$Sm$_\odot$) $^{\rm (b)}$}} 
\startdata
$^{80}$Kr  &  11.7 & 1 & 0.803  \\
$^{86}$Kr  & 27.0 & 2 & 0.142  \\
$^{86}$Sr  & 47.0 & 1 & 0.786  \\
$^{87}$Rb  & 35.3 & 2 & 0.181  \\
$^{88}$Sr  & 92.3 & 2 & 0.418  \\
$^{89}$Y  & 92.0 & 2 & 0.412  \\
$^{90}$Zr  & 72.2 & 3 & 0.905  \\
$^{96}$Zr  & 55.0 & 4 & 2.142  \\

\enddata                                                                 

\vspace{0.1em}
\hspace{9em}$^{(\rm a)}$ -- Arlandini et al.~(1999)
\vspace{0.1em}

\hspace{9em}$^{(\rm 1)}$ -- $^{80}$Kr and $^{86}$Sr, $s$-only isotopes. In this work we find an important contribution from $p$-process.
\vspace{0.1em}

\hspace{9em}$^{(\rm 2)}$ -- $^{86}$Kr, $^{87}$Rb, $^{88}$Sr, and $^{89}$Y are relics of the $s$-process seeds. 
\vspace{0.1em}

\hspace{9em}$^{(\rm 3)}$ -- $^{90}$Zr is a neutron magic nucleus at $N$=50. In this work we find an important contribution from $p$-process. 
\vspace{0.1em}

\hspace{9em}$^{(\rm 4)}$ -- $^{96}$Zr in our SN~Ia models gets an important contribution by neutron capture during $^{22}$Ne-burning.
\vspace{0.1em}

\vspace{0.1em}
\hspace{9em}$^{(\rm b)}$ -- Synthesized mass in DDT-a model normalized to $^{144}$Sm. For the results of this column, see Section~6.2.
\vspace{0.1em}

\end{deluxetable}

\begin{deluxetable}{cccccccccccc}
\tabletypesize{\scriptsize}
\rotate
\tablecaption{Synthesized Mass ($M_\odot$) in SN~Ia DDT-{\bf a} model (Figure~10, {\it upper panel})} 
\tablewidth{0pt}
\tablehead{
\colhead{Species} & \colhead{Abundance} &  \colhead{Species} & \colhead{Abundance} &
\colhead{Species} & \colhead{Abundance} &  \colhead{Species} & \colhead{Abundance} &  
\colhead{Species} & \colhead{Abundance} &  \colhead{Species} & \colhead{Abundance} 
} 
\startdata
$^{12}$C  &  1.5671D-02 & $^{50}$Cr & 5.4047D-04 &  $^{84}$Sr  & 4.5067D-07 &  $^{115}$Sn & 1.0004D-09 & $^{145}$Nd & 1.8189D-10 & $^{176}$Hf & 8.6456D-10 \\
$^{13}$C  &  9.2987D-08 & $^{52}$Cr & 1.1575D-02 &  $^{86}$Sr  & 1.8400D-06 &  $^{116}$Sn & 6.0745D-09 & $^{146}$Nd & 7.1862D-10 & $^{177}$Hf & 7.1442D-11  \\
$^{14}$N  &  2.1701D-05 & $^{53}$Cr & 1.5783D-03 &  $^{87}$Sr  & 1.4327D-07 &  $^{117}$Sn & 2.9810D-10 & $^{148}$Nd & 4.0872D-10 & $^{178}$Hf & 4.3106D-10  \\
$^{15}$N  &  2.1673D-08 & $^{54}$Cr & 6.6240D-06 &  $^{88}$Sr  & 8.3871D-06 &  $^{118}$Sn & 6.9873D-09 & $^{150}$Nd & 4.9146D-09 & $^{179}$Hf & 1.3962D-10  \\
$^{16}$O  &  1.4423D-01 & $^{55}$Mn & 1.4643D-02 &  $^{89}$Y   & 1.9993D-06 &  $^{119}$Sn & 6.6589D-10 & $^{144}$Sm & 1.3524D-08 & $^{180}$Hf & 1.5056D-09  \\
$^{17}$O  &  3.0220D-06 & $^{54}$Fe & 1.1991D-01 &  $^{90}$Zr  & 5.6021D-06 &  $^{120}$Sn & 7.5572D-09 & $^{147}$Sm & 1.4712D-10 & $^{180m}$Ta & 8.1889D-13  \\
$^{18}$O  &  2.5498D-08 & $^{56}$Fe & 5.8375D-01 &  $^{91}$Zr  & 7.3390D-08 &  $^{122}$Sn & 6.7889D-09 & $^{148}$Sm & 6.1696D-11 & $^{181}$Ta & 2.8885D-10  \\
$^{19}$F  &  7.1068D-10 & $^{57}$Fe & 1.5874D-02 &  $^{92}$Zr  & 2.8426D-07 &  $^{124}$Sn & 2.6818D-08 & $^{149}$Sm & 1.5191D-10 & $^{180}$W  & 9.6280D-10  \\
$^{20}$Ne &  5.6727D-03 & $^{58}$Fe & 4.0790D-05 &  $^{94}$Zr  & 2.3997D-07 &  $^{121}$Sb & 1.2773D-09 & $^{150}$Sm & 4.9988D-11 & $^{182}$W  & 8.4175D-10  \\
$^{21}$Ne &  1.8643D-06 & $^{59}$Co & 6.6528D-04 &  $^{96}$Zr  & 7.7244D-07 &  $^{123}$Sb & 1.9351D-09 & $^{152}$Sm & 1.3103D-10 & $^{183}$W  & 1.3810D-10  \\
$^{22}$Ne &  3.7666D-04 & $^{58}$Ni & 6.4124D-02 &  $^{93}$Nb  & 4.9745D-08 &  $^{120}$Te & 3.0647D-09 & $^{154}$Sm & 1.2263D-09 & $^{184}$W  & 1.8243D-10  \\
$^{23}$Na &  1.1737D-04 & $^{60}$Ni & 6.7264D-03 &  $^{92}$Mo  & 2.5027D-07 &  $^{122}$Te & 7.4450D-09 & $^{151}$Eu & 1.4435D-11 & $^{186}$W  & 1.3116D-09  \\
$^{24}$Mg &  2.0306D-02 & $^{61}$Ni & 7.7360D-05 &  $^{94}$Mo  & 1.6391D-07 &  $^{123}$Te & 4.2524D-10 & $^{153}$Eu & 1.3177D-10 & $^{185}$Re & 1.1175D-10  \\
$^{25}$Mg &  1.9054D-04 & $^{62}$Ni & 5.2097D-04 &  $^{95}$Mo  & 4.9595D-08 &  $^{124}$Te & 1.4025D-09 & $^{152}$Gd & 1.9977D-11 & $^{187}$Re & 2.2337D-10  \\
$^{26}$Mg &  1.9362D-04 & $^{64}$Ni & 4.6122D-06 &  $^{96}$Mo  & 3.8125D-08 &  $^{125}$Te & 2.7803D-10 & $^{154}$Gd & 5.7821D-11 & $^{184}$Os & 3.0456D-10  \\
$^{27}$Al &  8.2725D-04 & $^{63}$Cu & 1.5072D-05 &  $^{97}$Mo  & 9.1247D-09 &  $^{126}$Te & 1.9648D-09 & $^{155}$Gd & 1.3780D-11 & $^{186}$Os & 2.5331D-10  \\
$^{28}$Si &  3.3130D-01 & $^{65}$Cu & 6.5753D-06 &  $^{98}$Mo  & 3.2686D-08 &  $^{128}$Te & 2.1474D-09 & $^{156}$Gd & 9.6601D-11 & $^{187}$Os & 4.1744D-11  \\
$^{29}$Si &  1.1974D-03 & $^{64}$Zn & 6.4154D-06 &  $^{100}$Mo & 1.4678D-08 &  $^{130}$Te & 1.1202D-08 & $^{157}$Gd & 2.8455D-11 & $^{188}$Os & 1.3941D-10  \\
$^{30}$Si &  2.5425D-03 & $^{66}$Zn & 3.1710D-05 &  $^{96}$Ru  & 1.4367D-07 &  $^{127}$I  & 1.6683D-09 & $^{158}$Gd & 1.3021D-10 & $^{189}$Os & 3.3415D-11  \\
$^{31}$P  &  5.8732D-04 & $^{67}$Zn & 1.2094D-06 &  $^{98}$Ru  & 6.5865D-08 &  $^{124}$Xe & 8.2858D-09 & $^{160}$Gd & 7.4789D-10 & $^{190}$Os & 1.7001D-10  \\
$^{32}$S  &  1.3914D-01 & $^{68}$Zn & 2.9505D-06 &  $^{99}$Ru  & 1.3524D-08 &  $^{126}$Xe & 9.7036D-09 & $^{159}$Tb & 1.8889D-10 & $^{192}$Os & 1.4298D-09  \\
$^{33}$S  &  3.3104D-04 & $^{70}$Zn & 2.5358D-07 &  $^{100}$Ru & 1.6813D-08 &  $^{128}$Xe & 1.4251D-08 & $^{156}$Dy & 1.3456D-10 & $^{191}$Ir & 1.0247D-10  \\
$^{34}$S  &  3.5648D-03 & $^{69}$Ga & 1.1747D-06 &  $^{101}$Ru & 1.0647D-09 &  $^{129}$Xe & 6.7265D-10 & $^{158}$Dy & 1.2516D-10 & $^{193}$Ir & 3.6770D-10  \\
$^{36}$S  &  5.6228D-07 & $^{71}$Ga & 3.6326D-07 &  $^{102}$Ru & 1.4012D-09 &  $^{130}$Xe & 1.1922D-09 & $^{160}$Dy & 6.0440D-10 & $^{190}$Pt & 1.4588D-10  \\
$^{35}$Cl &  1.5530D-04 & $^{70}$Ge & 2.3483D-06 &  $^{104}$Ru & 7.8273D-09 &  $^{131}$Xe & 5.0030D-10 & $^{161}$Dy & 9.2518D-11 & $^{192}$Pt & 3.6798D-10  \\
$^{37}$Cl &  3.2379D-05 & $^{72}$Ge & 2.4812D-06 &  $^{103}$Rh & 1.3001D-09 &  $^{132}$Xe & 2.3573D-09 & $^{162}$Dy & 1.1898D-10 & $^{194}$Pt & 6.3110D-10  \\
$^{36}$Ar &  2.3840D-02 & $^{73}$Ge & 3.6625D-08 &  $^{102}$Pd & 1.8434D-08 &  $^{134}$Xe & 2.7812D-09 & $^{163}$Dy & 9.6694D-11 & $^{195}$Pt & 3.1544D-10  \\
$^{38}$Ar &  1.4908D-03 & $^{74}$Ge & 3.5906D-07 &  $^{104}$Pd & 4.8247D-09 &  $^{136}$Xe & 1.0643D-08 & $^{164}$Dy & 9.6427D-10 & $^{196}$Pt & 2.3987D-10  \\
$^{40}$Ar &  2.2358D-07 & $^{76}$Ge & 1.0567D-06 &  $^{105}$Pd & 6.7008D-10 &  $^{133}$Cs & 1.9744D-09 & $^{165}$Ho & 1.9165D-10 & $^{198}$Pt & 1.4554D-09  \\
$^{39}$K  &  9.8071D-05 & $^{75}$As & 1.9919D-07 &  $^{106}$Pd & 1.4563D-09 &  $^{130}$Ba & 8.1235D-09 & $^{162}$Er & 2.8935D-10 & $^{197}$Au & 6.8591D-10  \\
$^{40}$K  &  8.9145D-08 & $^{74}$Se & 4.0425D-07 &  $^{108}$Pd & 1.4541D-09 &  $^{132}$Ba & 5.8825D-09 & $^{164}$Er & 9.0568D-10 & $^{196}$Hg & 1.8911D-09  \\
$^{41}$K  &  5.9740D-06 & $^{76}$Se & 8.9696D-07 &  $^{110}$Pd & 1.0164D-08 &  $^{134}$Ba & 4.9567D-09 & $^{166}$Er & 5.6222D-10 & $^{198}$Hg & 2.4552D-09  \\
$^{40}$Ca &  2.0560D-02 & $^{77}$Se & 7.6909D-09 &  $^{107}$Ag & 5.1444D-10 &  $^{135}$Ba & 1.1355D-09 & $^{167}$Er & 8.5695D-11 & $^{199}$Hg & 3.5610D-10  \\
$^{42}$Ca &  3.3420D-05 & $^{78}$Se & 3.4351D-07 &  $^{109}$Ag & 1.4341D-09 &  $^{136}$Ba & 5.1733D-09 & $^{168}$Er & 1.2140D-10 & $^{200}$Hg & 2.4156D-09  \\
$^{43}$Ca &  2.1826D-07 & $^{80}$Se & 2.1413D-07 &  $^{106}$Cd & 2.7891D-08 &  $^{137}$Ba & 1.7809D-09 & $^{170}$Er & 7.1327D-10 & $^{201}$Hg & 5.2630D-10  \\
$^{44}$Ca &  1.5950D-05 & $^{82}$Se & 8.2393D-07 &  $^{108}$Cd & 7.2676D-09 &  $^{138}$Ba & 7.1886D-08 & $^{169}$Tm & 2.5781D-10 & $^{202}$Hg & 2.2432D-09   \\
$^{46}$Ca &  8.8764D-08 & $^{79}$Br & 4.8509D-08 &  $^{110}$Cd & 2.3744D-08 &  $^{138}$La & 2.7045D-11 & $^{168}$Yb & 8.5902D-10 & $^{204}$Hg & 3.5097D-09    \\
$^{48}$Ca &  1.5001D-08 & $^{81}$Br & 2.1398D-07 &  $^{111}$Cd & 8.9753D-10 &  $^{139}$La & 6.6975D-09 & $^{170}$Yb & 1.1057D-09 & $^{203}$Tl & 8.4790D-10    \\
$^{45}$Sc &  4.6983D-07 & $^{78}$Kr & 1.0037D-07 &  $^{112}$Cd & 2.1349D-09 &  $^{136}$Ce & 1.4674D-09 & $^{171}$Yb & 1.0724D-10 & $^{205}$Tl & 2.9824D-09    \\
$^{46}$Ti &  1.9035D-05 & $^{80}$Kr & 7.5378D-07 &  $^{113}$Cd & 1.4926D-10 &  $^{138}$Ce & 2.5269D-09 & $^{172}$Yb & 5.9715D-10 & $^{204}$Pb & 1.9311D-09    \\
$^{47}$Ti &  1.2317D-06 & $^{82}$Kr & 3.7126D-07 &  $^{114}$Cd & 2.3586D-09 &  $^{140}$Ce & 1.6945D-08 & $^{173}$Yb & 1.2626D-10 & $^{206}$Pb & 4.8271D-09   \\
$^{48}$Ti &  4.4708D-04 & $^{83}$Kr & 2.1573D-08 &  $^{116}$Cd & 1.3732D-08 &  $^{142}$Ce & 7.8332D-09 & $^{174}$Yb & 3.5181D-10 & $^{207}$Pb & 6.0427D-09   \\
$^{49}$Ti &  3.6257D-05 & $^{84}$Kr & 2.9946D-07 &  $^{113}$In & 1.6635D-09 &  $^{141}$Pr & 2.7194D-09 & $^{176}$Yb & 1.3185D-09 & $^{208}$Pb & 3.3585D-08   \\
$^{50}$Ti &  2.3077D-07 & $^{86}$Kr & 1.1256D-06 &  $^{115}$In & 1.2077D-09 &  $^{142}$Nd & 1.2709D-08 & $^{175}$Lu & 1.8984D-10 & $^{209}$Bi & 1.8919D-10   \\
$^{50}$V  &  7.8373D-08 & $^{85}$Rb & 2.4553D-07 &  $^{112}$Sn & 3.6123D-08 &  $^{143}$Nd & 7.0663D-10 & $^{176}$Lu & 1.1738D-10 &            &              \\
$^{51}$V  &  1.2992D-04 & $^{87}$Rb & 3.8945D-07 &  $^{114}$Sn & 1.1931D-08 &  $^{144}$Nd & 1.0278D-09 & $^{174}$Hf & 6.2584D-10 &            &              \\
\enddata                                                                 
\end{deluxetable}

\end{document}